\documentclass[a4paper,fleqn,usenatbib]{mnras}
\usepackage{lipsum} 
\usepackage[T1]{fontenc}
\usepackage{ae,aecompl}

\usepackage{graphicx}	
\usepackage{amsmath}	
\usepackage{amssymb}	
\usepackage{subcaption} 
\title[Bulges in galaxy groups]{Pseudo bulges in galaxy groups: the role of environment in secular evolution}

\author[P. K. Mishra et al.]{
Preetish K. Mishra,$^{1}$\thanks{E-mail: preetish@ncra.tifr.res.in}
Yogesh Wadadekar,$^{1}$\thanks{E-mail: yogesh@ncra.tifr.res.in}
Sudhanshu Barway$^{2}$\thanks{E-mail: barway@saao.ac.za }
\\
$^{1}$National Centre for Radio Astrophysics, TIFR, Post Bag 3, Ganeshkhind, Pune 411007, India\\
$^{2}$South African Astronomical Observatory, PO Box 9, 7935, Observatory, Cape Town, South Africa\\
}

\date{to appear in the MNRAS.}

\pubyear{2016}

\begin{document}
\label{firstpage}
\pagerange{\pageref{firstpage}--\pageref{lastpage}}
\maketitle

\begin{abstract}
We examine the dependence of the fraction of galaxies containing pseudo bulges on environment for a flux limited sample of $\sim$5000 SDSS galaxies. We have separated bulges into classical and pseudo bulge categories based on their position on the Kormendy diagram. Pseudo bulges are thought to be formed by internal processes and are a result of secular evolution in galaxies. We attempt to understand the dependence of secular evolution on environment and morphology. Dividing our sample of disc+bulge galaxies based on group membership into three categories: central and satellite galaxies in groups and isolated field galaxies, we find that pseudo bulge fraction is almost equal for satellite and field galaxies. Fraction of pseudo bulge hosts in central galaxies is almost half of the fraction of pseudo bulges in satellite and field galaxies. This trend is also valid when only galaxies are considered only spirals or S0. Using the projected fifth nearest neighbour density as measure of local environment, we look for the dependence of pseudo bulge fraction on environmental density. Satellite and field galaxies show very weak or no dependence of pseudo bulge fraction on environment. However, fraction of pseudo bulges hosted by central galaxies decreases with increase in local environmental density. We do not find any dependence of pseudo bulge luminosity on environment. Our results suggest that the processes that differentiate the bulge types are a function of environment while processes responsible for the formation of pseudo bulges seem to be independent of environment. 
\end{abstract}




\section{Introduction}
 
Recent progress in our understanding of the central component of disc galaxies has expanded our knowledge of galaxy formation and evolution. We now know that the central component i.e. the bulge,
comes in two flavours. Classical bulges are thought to be formed by mergers \citep{Aguerri2001} or sinking of giant gas clumps found in high redshift discs to central region of the galaxy and formation of these bulges through violent relaxation and starbursts \citep{Dekel2009,Cacciato2012,Forbes2014,Kormendy2016}. Pseudo bulges on the other hand are thought to be the product of internal processes and have secularly evolved through time \citep{Kormendy&Kennicutt2004}. Difference in the formation scenario for these two bulge types makes them fundamentally different from one another, which is also reflected in their distinct properties. Pseudo bulges exhibit nuclear morphological features which are characteristic of galaxy discs such as a nuclear bar, spiral or ring \citep{Carollo1998} while classical bulges are featureless. Also, pseudo bulges are composed of a younger stellar population with a flatter radial velocity dispersion profile as compared to that of classical bulges \citep{Gadotti2009,Fabricius2012}. The two types of bulges behave differently with respect to several well known correlations between structural parameters of galaxies. For example, \citet{Gadotti2009} has shown that the two bulge types occupy different regions of the parameter space, when plotted as different projections of the fundamental plane \citep{Djorgovski1987}. A smooth transition from one type of bulge to another as seen in these correlations also points to possible existence of composite bulges with mixed properties \citep{Gadotti2009, Fisher2010}. Properties of composite bulges have been explored in detail in recent works for e.g. \citet{Erwin2015} but much work is needed to put strong limits on the frequency of occurrence of composite bulges in galaxies.\par

Most previous work on bulges \citep{Carollo1998,Fisher2008,Gadotti2009,Fabricius2012} has focussed on the relation between bulges and properties of their host galaxies, which in turn, have been used as criteria for classification of bulge type. As a result, there has been significant increase in our understanding of the importance of bulges in the evolution of their host galaxies and associated black holes, AGN etc. \citep{Gadotti2009a,Kormendy2011,Lorenzo2014,Vaghmare2015}. At this time there exist a number of simulations which are successful in forming classical bulges via mergers \citep{Aguerri2001} or via clump instabilities \citep{Elmegreen2008}. Also, our understanding of secular evolution is now detailed enough to qualitatively explain commonly occurring morphological features in galaxies such as nuclear rings, nuclear bars and pseudo bulges. It has been shown in simulations that bar driven secular evolution can form inner and outer rings, pseudo bulges and structure that resemble nuclear spirals as observed in disc galaxies \citep{Simkin1980,Sanders1980,Athanassoula1992,Salo1999,Rautiainen2000}. On the other hand, studies attempting to quantitatively understand the process of secular evolution in diverse environments are not very common in the literature. \par

A study of the Virgo cluster by \citet{Kormendy2009} shows that 2/3 of stellar mass is in elliptical galaxies alone. Information on other extreme of environmental density regime comes from \citet{Fisher2011}. By studying galaxies within the local 11 Mpc volume in low density regions, they report that 1/4 of stellar mass is contained within ellipticals and classical bulges while rest 3/4 of mass is distributed in pseudo bulges, discs and bulge-less galaxies. These two observations have led the authors to conclude that the process driving distribution of bulge type appears to be a strong function of environment. There are only a few works which explore the effect of environment specifically on bulges. \citet{Hudson2010} have studied the colour of bulges and discs in clusters and found that the bulge colour does not depend on environment. \citet{Lackner2013} distinguished between galaxies having a quiescent bulge and a star forming bulge based on the strength of the 4000 {\AA} break ($D_n(4000)$ index). They have associated quiescent bulges with classical bulges and star forming bulges with pseudo bulges. In their work, the classical bulge profile is modelled as having de Vaucouleurs profile with bulge S\'ersic index $n_b = 4$ and pseudo bulges are modelled with an exponential profile with $n_b = 1$. Using fifth projected neighbourhood density as a measure of local environment, they show a strong increase in fraction of galaxies hosting a classical bulge with increase in local density. On the other hand, fraction of galaxies hosting a pseudo bulge decreases slowly as one goes from a lower to a higher local density environment. \citet{Lackner2013} have focussed on studying the dependence of galaxy properties of classical bulges on the environment. There is a need to explore properties of pseudo bulges over a wide range of environmental density in order to expand our knowledge of secular evolution and make it more quantitative. Dependence of distribution of pseudo bulges and their intrinsic properties on environment will help us understand how environment affects the processes that govern distribution and formation of pseudo bulges. \par

 In this work, we have explored the dependence of bulge type on
 environment as well as on galaxy morphology. Our sample spans a wide
 range of environmental density and is composed of mainly
 isolated/field galaxies and galaxy groups with available
 morphological information for each object. We have identified and
 classified our sample of S0 and spiral galaxies into classical and
 pseudo bulge host galaxies. We further divide our sample by
 galaxy group association, into three categories: 1. field galaxies not
 belonging to any group, 2. galaxies which reside in the center of galaxy
 groups and 3. their satellite galaxies. We investigate the dependence of
 bulge type on environments in these three categories. The paper is
 organised as follows. Section 2 describes the data and sample
 selection. Section 3 describes our results and Section 4 summarizes
 the findings and the implications. Throughout this work, we have used
 the WMAP9 cosmological parameters : $H_0$ = 69.3 km
 s$^{-1}$Mpc$^{-1}$, $\Omega_{m}$ = 0.287 and $\Omega_{\Lambda}$=
 0.713. Unless otherwise noted, photometric measurements used are in the
 SDSS $r$ band. All logarithms are to base 10.
 
 \section{Data and sample selection}
 
To aim for a sample suitable for studying bulges in galaxy groups, we
started with data from \citet{Meert2015} which provides a catalogue of
nearly 700,000 spectrocopically selected galaxies drawn from the SDSS
DR7 in SDSS $g$, $r$ and $i$ bands \citep{Meert2016}. This catalogue is a
flux limited sample with $r$-band Petrosian magnitude for all galaxies
having magnitude in range $14<r<17.7$ and provides 2D, PSF corrected
de Vaucouleurs, S\'ersic, de Vaucouleurs+Exponential and
S\'ersic+Exponential fits of galaxies with flags indicating goodness of
the fit, using the PyMorph pipeline \citep{Vikram2010}. \par

We cross matched the \citet{Meert2015} catalogue with the data
provided in \citet{Nair2010} which is a catalogue of detailed visual
classification for nearly 14,000 spectroscopically targeted galaxies
in the SDSS DR4. The \citet{Nair2010} catalogue is a flux limited sample
with an extinction corrected limit of $g<16$ mag in SDSS $g$ band,
spanning the redshift range $0.01 < z < 0.1$. In addition to
morphological T-type classification it also provides measurements of
stellar mass of each object taken from \citet{Kauffmann2003} which used stellar absorption-line indices and broadband photometry from SDSS to estimate the stellar mass of galaxies. The \citet{Nair2010} catalogue also contains information on
average environmental density from \citet{Baldry2006} and information
on galaxy groups from \citet{Yang2007} such as group mass, group
luminosity, group halo mass, group richness etc. We have made use of
all relevant information on individual galaxies and groups as provided
in \citet{Nair2010} in our work.  Our cross match of these two
catalogues resulted in a sample of 8929 galaxies on which we have
further imposed condition a requirement of a ``good fit'' as given in \citet{Meert2015} and
which is described below.\par

To obtain our final sample of galaxies we have first removed galaxies with bad
fits and problematic two component fits as indicated by flags in
\citet{Meert2015}. The three categories of good fits in this catalogue
are as follows. \\ \\ (i) Good two component fits : which includes
galaxies where both bulge and disc components have reliable estimates
\\ (ii) Good bulge fits : which includes galaxies where disc estimates
are unreliable while the bulge measurements are trustworthy \\ (iii)
Good disc fits : which includes galaxies where bulge estimates are
unreliable but the disc measurements are trustworthy.\\ \\ Since the
focus of this study is on bulges we have retained galaxies in the first two categories
and have discarded those in the third one. Additional constraints comes from
fact that two component fits with bulge S\'ersic index $n \geq 8$ can be used
for total magnitude and radius measurement but they have unreliable
subcomponents. We have taken a conservative approach and have retained
only the galaxies having bulge S\'ersic index $n < 8$. \par

After applying all selection criteria mentioned above, we are left with
4991 galaxies , 2026 of which are spirals ($1\leq T \leq 9$), 1732 are
S0s ($-3\leq T \leq 0$) and 1233 are ellipticals ($-4\leq T$). From
here onwards, we will collectively refer to the population of
spirals+S0s as disc galaxies. Out of the 3758 disc galaxies in our sample,
we have information about the group properties of 3641 galaxies from \citet{Yang2007}.


\begin{figure*}
  \begin{subfigure}[b]{0.35\textwidth}
      \includegraphics[width=\textwidth]{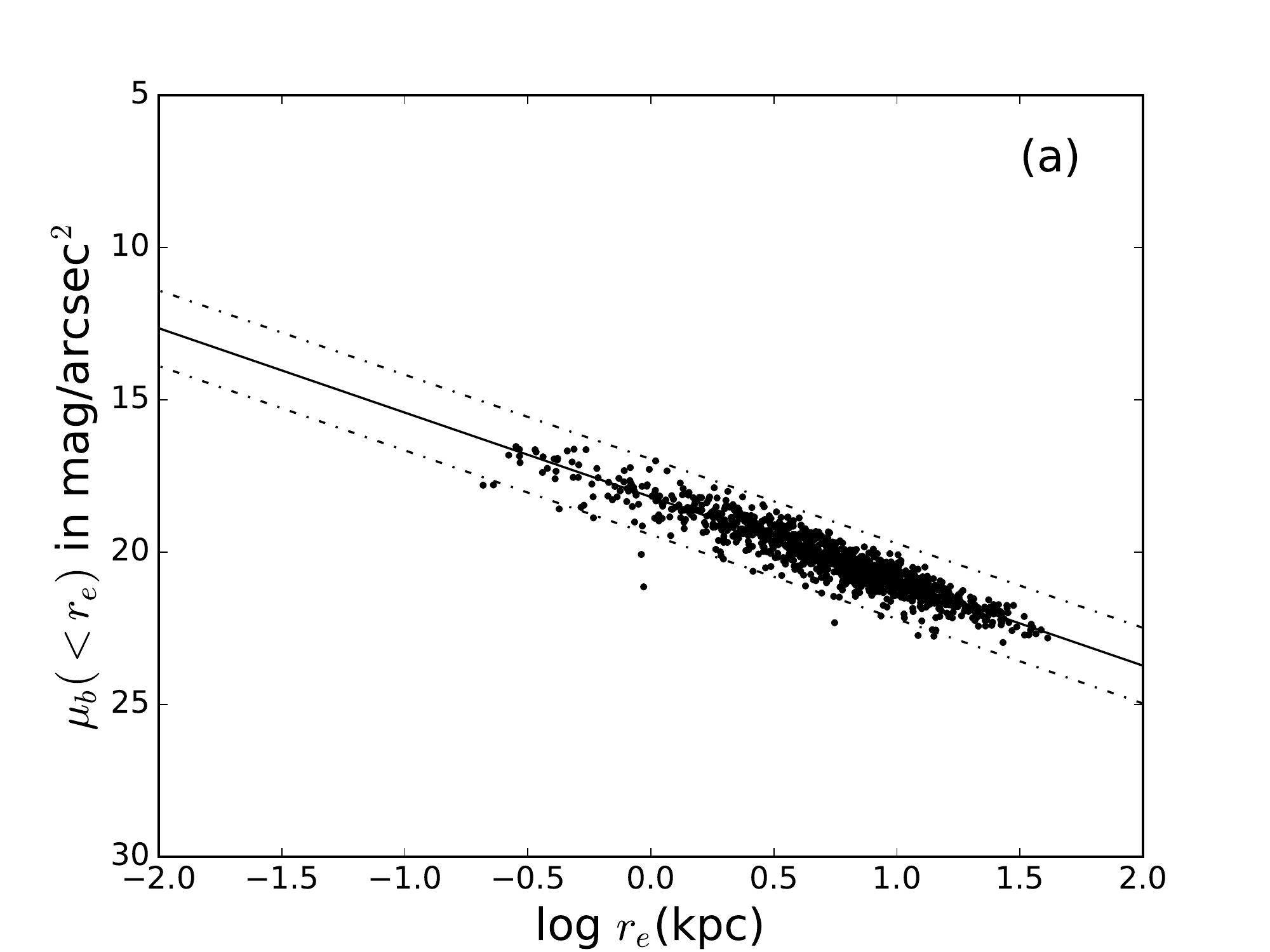}
    \phantomcaption
    \label{fig:elliptical}
  \end{subfigure}
  \hspace{-1.9em}%
  \begin{subfigure}[b]{0.35\textwidth}
    \includegraphics[width=\textwidth]{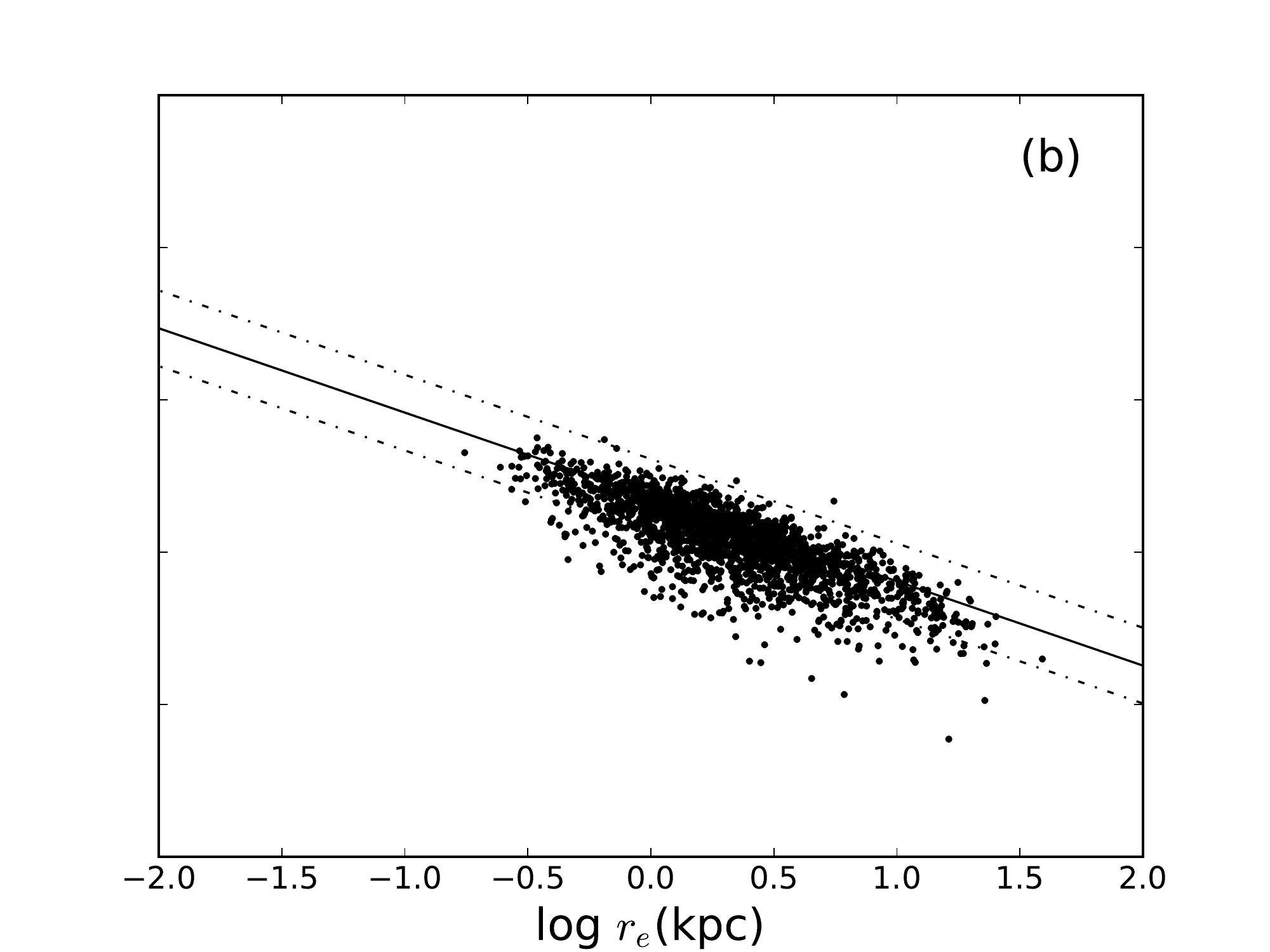}
    \phantomcaption
    \label{fig:spiral}
  \end{subfigure}
  \hspace{-1.9em}%
  \begin{subfigure}[b]{0.35\textwidth}
    \includegraphics[width=\textwidth]{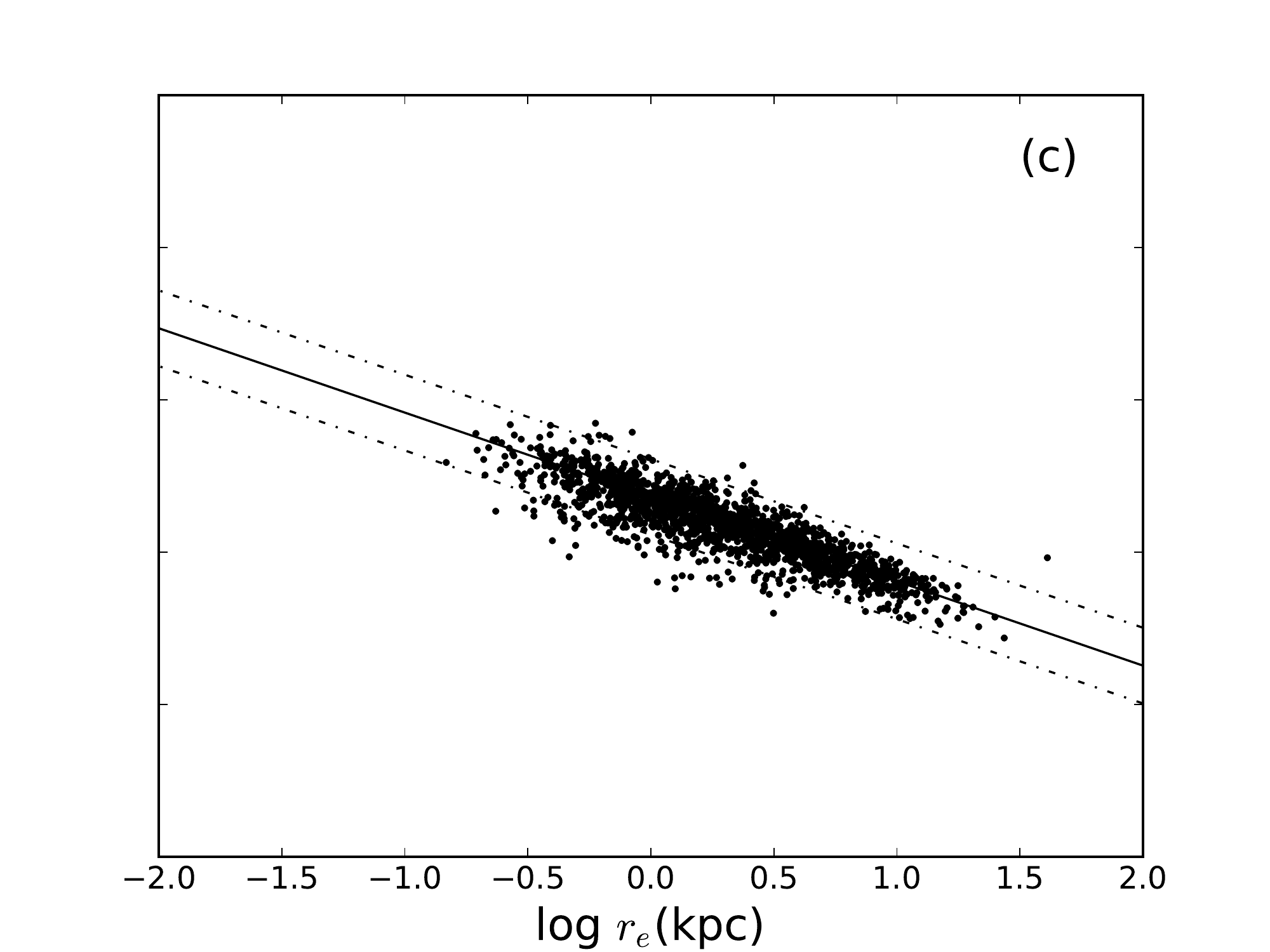}
    \phantomcaption
    \label{fig:S0}
  \end{subfigure}
\label{fig:Kormendy}
\vspace{-1.5em}
\caption{(a) Kormendy relation for elliptical galaxies. The solid
line is best fit line to these ellipticals while the two dashed lines
enclose 3$\sigma$ scatter from best fit; (b) bulges of spiral galaxies; (c) bulges of S0 galaxies in our sample}  
\end{figure*}

\begin{figure*}
  \begin{subfigure}[b]{0.4\textwidth}
    \includegraphics[width=\textwidth]{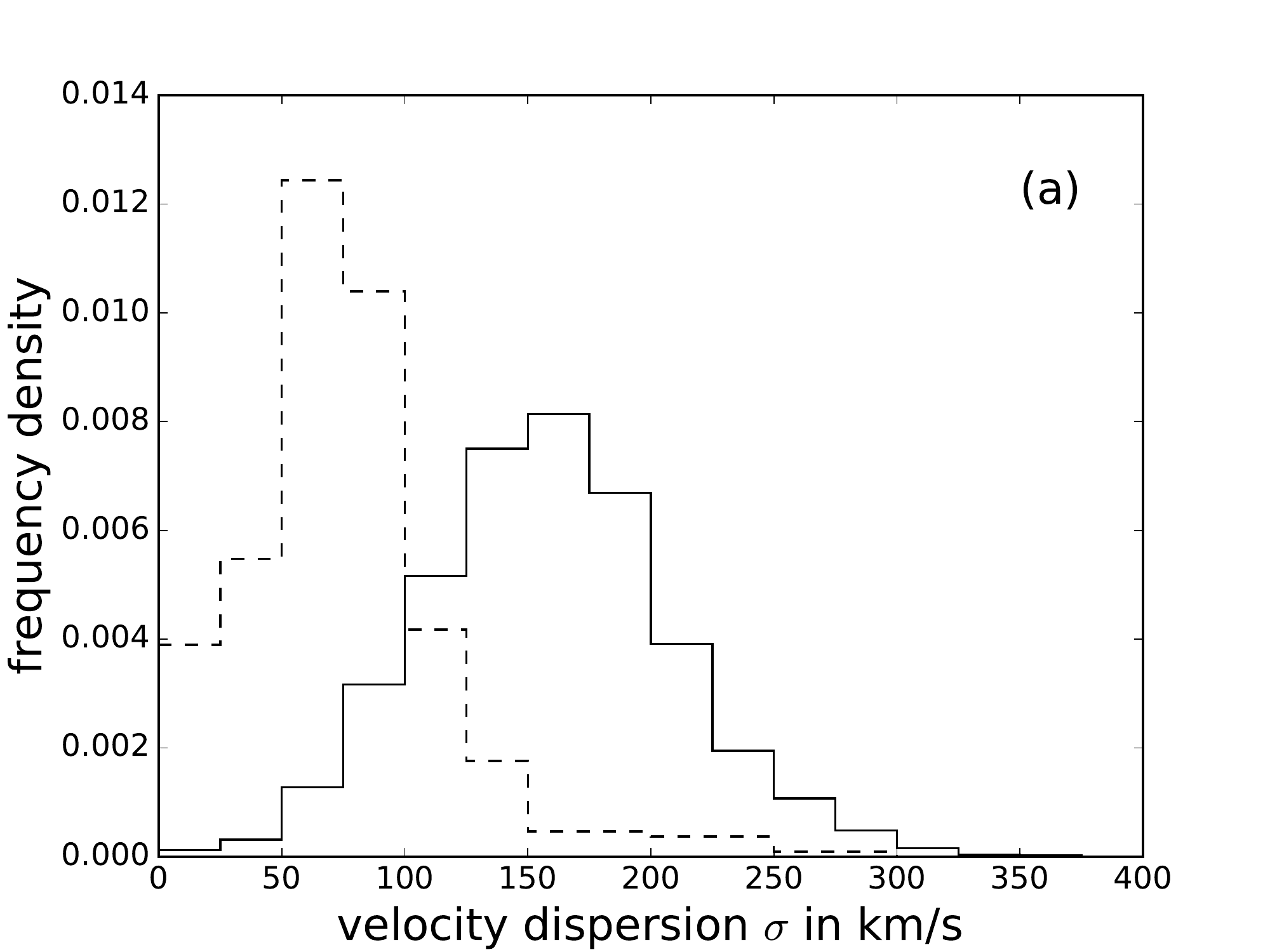}
    
    \phantomcaption
    \label{fig:velocity}
    
  \end{subfigure}
  \begin{subfigure}[b]{0.4\textwidth}
    \includegraphics[width=\textwidth]{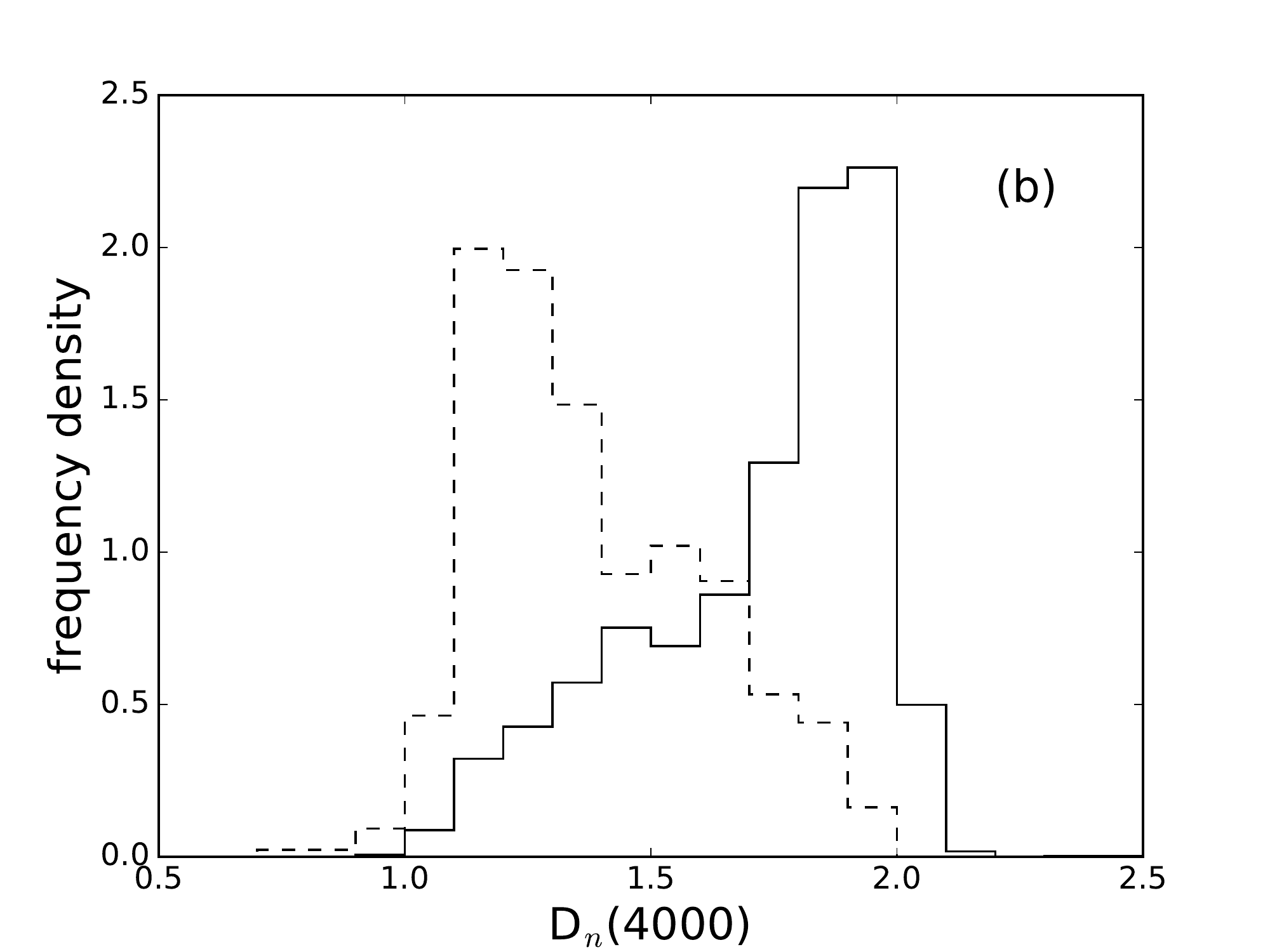}
    
    \phantomcaption
    \label{fig:dn4000}
    
  \end{subfigure}
  
\caption{ Distribution of (a) central velocity dispersion and (b) D$_n$(4000) index for classical and pseudo bulges in our sample. Both distributions have been normalised by area. Solid and dashed lines denote classical and pseudo bulge host galaxies respectively.}  
\end{figure*}

\begin{figure}
   \includegraphics[width=\columnwidth]{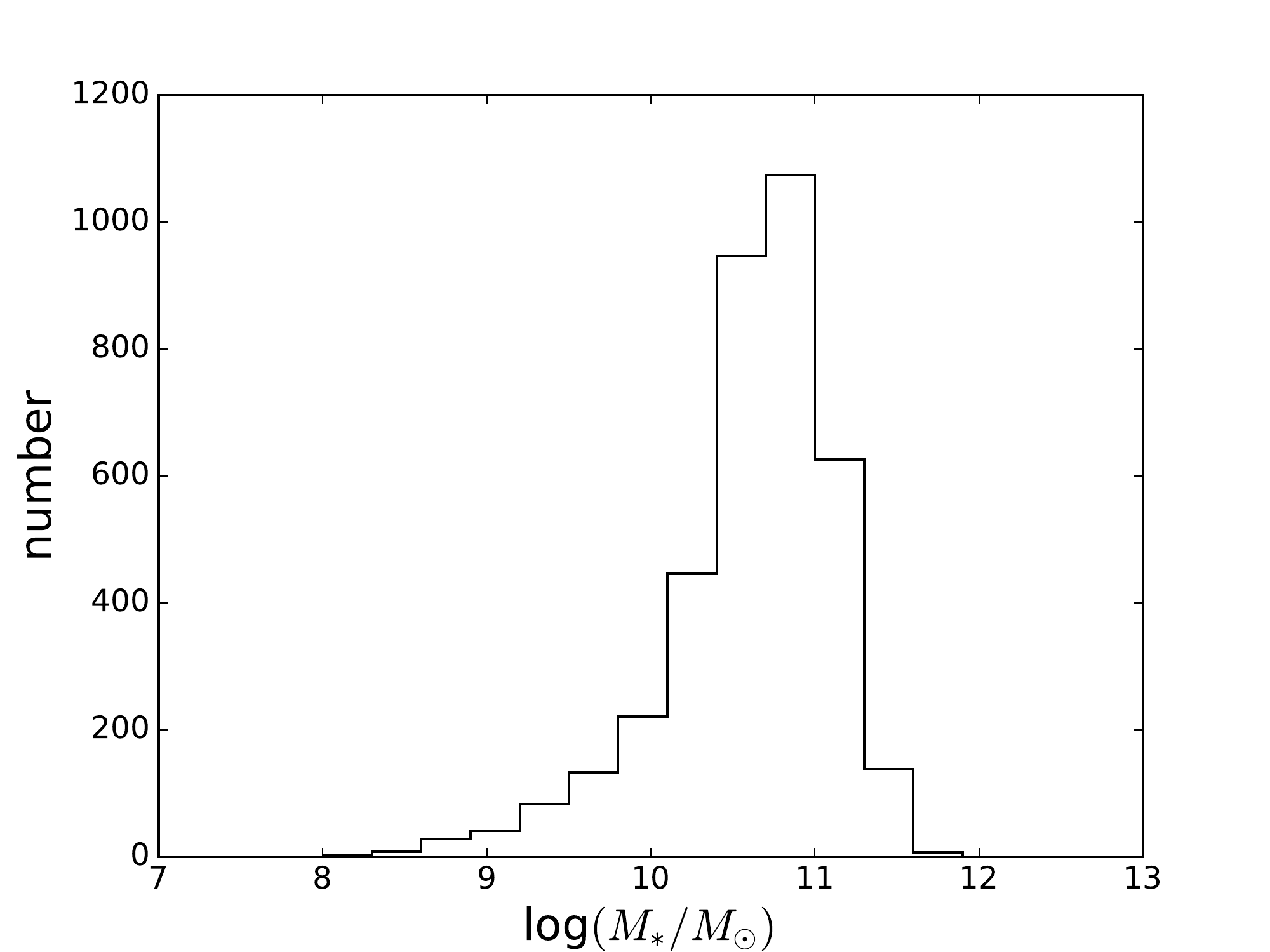}
   \caption{Stellar mass distribution of disk galaxies in our sample}
   \label{fig:mass}
\end{figure}

\begin{table}
	\centering
	
	\caption{Number of classical and pseudo bulges in spiral and S0 galaxies in our sample}
	\label{tab:all_all}
	\resizebox{\columnwidth}{!}{%
	\begin{tabular}{lccr} 
		\hline
		Bulge type & Disc galaxies& S0  & Spiral \\
		\hline
		
		All bulges & 3758 & 1732 &2026\\
		Classical bulge & 3327 & 1639  &1688\\
		Pseudo bulge & 431 & 93 &338\\
		Pseudo bulge fraction(\%) & 11.47 & 5.37 & 16.68 \\
		\hline
	\end{tabular}
	}
\end{table}

\begin{table}
	\centering
	\caption{Distribution of elliptical galaxies in our sample into central, satellite and group categories.}
	\label{tab:elliptical_table}
	
	\begin{tabular}{lccr} 
		\hline
		All ellipticals  & Central & Satellite & field \\
		\hline
		
		1233 & 684 & 222 & 290\\
		
		\hline
	\end{tabular}
	
\end{table}

\begin{table}
	\centering
	
	\caption{Number of classical and pseudo bulges in all disc galaxies classified as central, satellite and field galaxies.}
	\label{tab:disk_all}
	\resizebox{\columnwidth}{!}{%
	\begin{tabular}{lccr} 
		\hline
		Bulge type & Central& Satellite & Field \\
		\hline
		
		All bulges & 1119 & 752 &1770\\
		Classical bulge & 1058 & 657  &1516\\
		Pseudo bulge & 61 & 95 &254\\
		Pseudo bulge fraction(\%) & 5.45 & 12.63 & 14.35 \\
		Median galaxy stellar mass (log$M_{\odot}$) & 10.865 & 10.624 & 10.627 \\
		\hline
	\end{tabular}
	}
\end{table}

\begin{table}
	\centering
	\caption{Number of classical and pseudo bulges in all S0 galaxies classified as central, satellite and field galaxies. }
	\label{tab:S0_all}
	\resizebox{\columnwidth}{!}{%
	\begin{tabular}{lccr} 
		\hline
		Bulge type & Central& Satellite & Field \\
		\hline
		
		All bulges & 555 & 374 & 741\\
		Classical bulge & 541 & 353  & 690\\
		Pseudo bulge & 14 & 21 & 51\\
		Pseudo bulge fraction(\%) & 2.52 & 5.61 & 6.88 \\
		Median galaxy stellar mass (log$M_{\odot}$) & 10.854 & 10.617 & 10.605 \\
		\hline
	\end{tabular}
	}
\end{table}

\begin{table}
	\centering
	\caption{Number of classical and pseudo bulges in all spiral galaxies classified as central, satellite and field galaxies.) }
	\label{tab:spiral_all}
	\resizebox{\columnwidth}{!}{%
	\begin{tabular}{lccr} 
		\hline
		Bulge type & Central& Satellite & Field \\
		\hline
		
		All bulges & 564 & 378 &1029\\
		Classical bulge & 517 & 304 &826\\
		Pseudo bulge & 47 & 74 &203\\
		Pseudo bulge fraction(\%) & 8.33 & 19.58 & 19.73 \\
		Median galaxy stellar mass (log$M_{\odot}$) & 10.877 & 10.629 & 10.651 \\
		\hline
	\end{tabular}
	}
\end{table}

\section{Results}
\subsection{Identifying pseudo bulges}
A common practice in studies of bulges is to classify them on basis of the S\'ersic index. In this method, pseudo bulges are defined as those having S\'ersic index below a certain threshold. Usually this threshold value is taken to be $n = 2$ \citep{Ho2014,Ribeiro2016}. However, measurement of S\'ersic index from ground based telescopes are reported to have errors as large as 0.5 \citep{Gadotti2008,Durbala2008}. Also, S\'ersic index $n$ and effective radius ($r_e$) have degenerate errors which leads to additional error in $n$ due to uncertainty in measurement in $r_e$. Hence, using a specific S\'ersic index threshold for bulge classification may lead to ambiguity. Therefore, we have refrained from using the S\'ersic index to classify bulges in favour of a better physically motivated classification criteria due to \citet{Gadotti2009} which has been used in recent works for eg. \citet{Vaghmare2013}. \par 

This criteria involves classification of bulge types based on their position on the Kormendy diagram \citep{Kormendy1977}. The Kormendy diagram is a plot of the average surface brightness of the bulge within its effective radius ($\mu_b(< r_e)$) against logarithm of effective radius
$r_e$. Elliptical galaxies are known to obey a tight linear relation on this diagram. Classical bulges are thought to be structurally similar to ellipticals and therefore obey a similar relation while pseudo bulges being structurally  different, lie away from it. Any bulge that deviates more that three times the r.m.s. scatter from the best fit relation for ellipticals is classified as
pseudo bulge by this criterion \citep{Gadotti2009}.  \par 

The Kormendy diagram for our sample is shown in Figure \ref{fig:elliptical}. The best fit line was obtained by plotting elliptical galaxies using $r$ band data. The equation of the best fit line is \\ \\ $\langle \mu_b(<r_e)\rangle$ = (2.768 $\pm$ 0.0306)log$r_e$ + 18.172 $\pm$ 0.0255\\ \\  

The rms scatter in $\langle \mu_b(<r_e) \rangle$ around the best fit line is 0.414. The dashed lines encloses region of 3 times rms scatter from the fit. All galaxies lying outside the region enclosed by the dashed lines are taken to be pseudo bulges. \par

We have separated the disk galaxies having S0 and spiral morphology and have plotted them on the Kormendy diagram as shown in Figures \ref{fig:spiral} and \ref{fig:S0} respectively. It is clear from these figures that the number of pseudo bulge host galaxies are higher in spirals than in S0 galaxies. We have found that out of 2026 spiral galaxies, 338 (16.68 percent of spiral population) host a pseudo bulge. On the other hand only 93 (5.37 percent of S0s) out of a total of 1732 number of S0 galaxies are pseudo bulge hosts. This result is summarised in Table \ref{tab:all_all}.

To test the robustness of our classification criterion, we have compared the classical and pseudo bulges in our sample with respect to the properties in which two bulge types are expected to show different behaviour. Secularly evolved pseudo bulges, due to their disk like stellar kinematics are dynamically colder systems compared to the merger generated classical bulges. This property is reflected in the different values of velocity dispersion of the two bulge types. For eg. \citet{Fisher2016} have shown that on an average pseudo bulges have lower central velocity dispersion than classical bulges. Classifying the bulge type based on the S\'ersic index they have seen that pseudo bulges have average velocity dispersion $\sim$90 km/s whereas this value is $\sim$160 km/s for classical bulges. We have obtained the values of central velocity dispersion for the galaxies in our sample from \citet{Nair2010}. After applying aperture correction, we have plotted the distribution of central velocity dispersion for classical and pseudo bulge host galaxies in our sample which is shown in Figure \ref{fig:velocity}. One can see a bimodal distribution of central velocity distribution with respect to the bulge type. Pseudo bulges are found to have a lower value of central velocity dispersion, with their distribution peaking around $\sim$60 km/s in contrast to the distribution of the same for classical bulges which peaks around $\sim$160 km/s, in agreement with expected trends.\par 

Previous studies \citep{Gadotti2001,Fisher2006} have also indicated that pseudo bulges exhibit star forming activity as opposed to classical bulges which are mainly composed of old stars. To separate old and young stellar population in galaxies, we have used the strength of the 4000 {\AA} spectral break which arises due to accumulation of absorption lines of mainly metals in the atmosphere of old, low mass stars and by a lack of hot, blue stars in galaxies. The strength of this break is quantified by D$_n$(4000) index. In literature, one can find several definition of this index which are available. In this work, we have used the definition of this index as provided in \citet{Balogh1999}. A low value of D$_n$(4000) index denotes young stellar population. We have taken the D$_n$(4000) index measurement from SDSS DR7 MPA/JHU catalogue\footnote{\url{http://wwwmpa.mpa-garching.mpg.de/SDSS/DR7/}} and have plotted it's distribution for classical and pseudo bulges in our sample as shown in Figure \ref{fig:dn4000}.
As expected, our pseudo bulges have lower value of the D$_n$(4000) index which peaks around value $\sim$1.2 as compared to  the classical bulges peaking around $\sim$1.8. A similar bimodal distribution of D$_n$(4000) index with respect to bulge types, has also been found in works \citep{Fisher2016} that employ only bulge S\'ersic index cutoff $n=2$ to classify bulges.  \citet{Gadotti2009} have compared this bimodal distribution of D$_n$4000 when bulges are classified using the Kormendy relation only and when classification is based on threshold bulge S\'ersic index. They find that peaks of distribution of D$_n$(4000) for classical and pseudo bulge are closer when bulges are identified using the S\'ersic index as compared to distance between peaks when the Kormendy relation is used for classification. Our result for distribution of D$_n$(4000) is consistent with trend reported in \citet{Gadotti2009} which uses the Kormendy diagram for bulge classification. This gives support to the bulge classification criteria that we have used. \par

At this point, we would like to mention that our sample is a flux limited sample with an extinction corrected flux limit of $g<16$ mag in the SDSS $g$ band which makes it biased towards bright and massive galaxies. In Figure \ref{fig:mass} we have plotted the stellar mass distribution of disk galaxies in our sample. Its clear that our sample is biased towards massive galaxies.
As will be seen in later part of this paper that pseudo bulges are more common in galaxies having low stellar mass. Hence, the result presented here on fraction of pseudo bulges is applicable only to bright and massive galaxies.


\begin{figure}
   \includegraphics[width=\columnwidth]{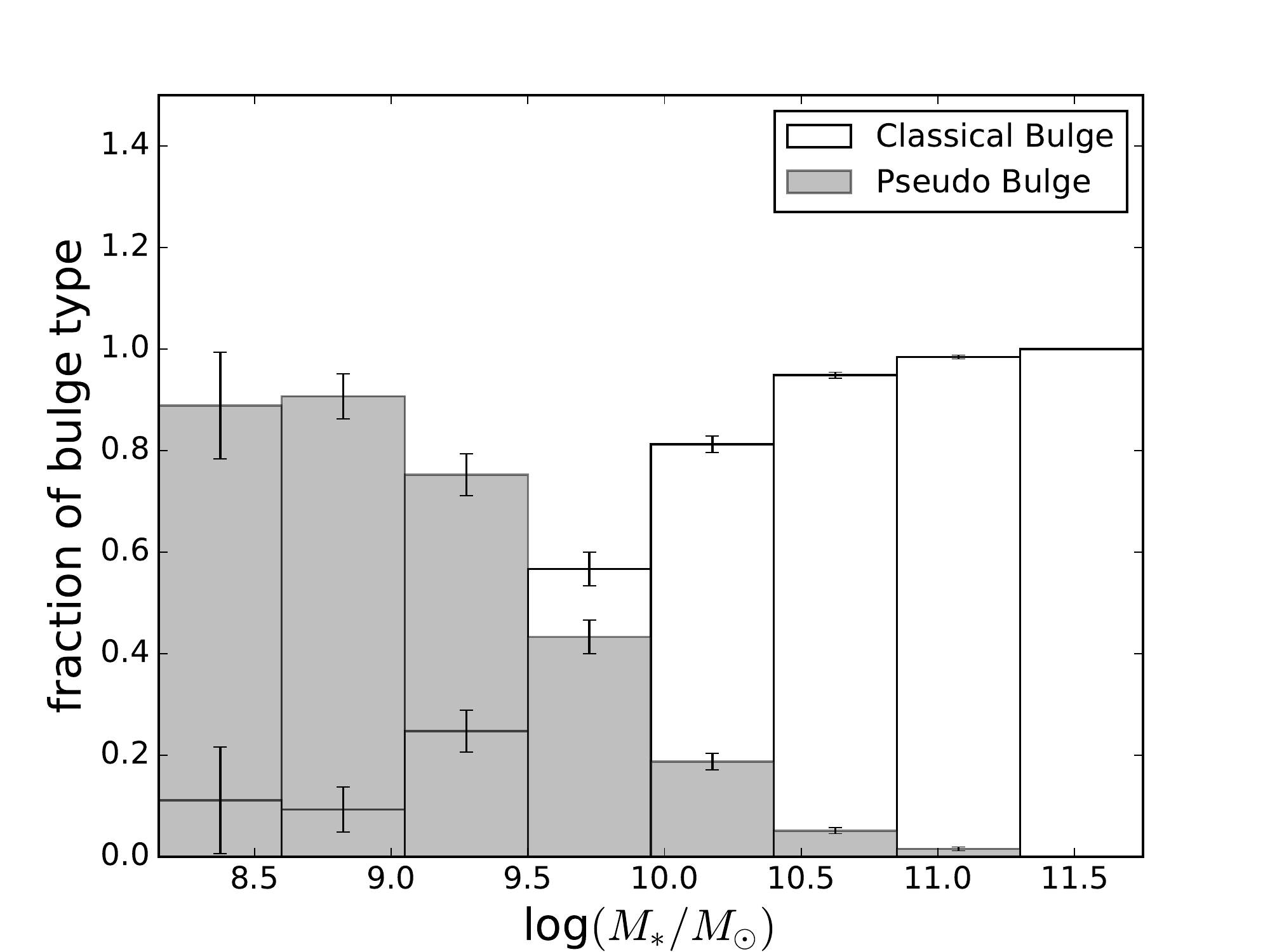}
   \caption{Dependence of classical/pseudo bulge fraction on host galaxy stellar mass. Grey shaded region denoting pseudo bulge is placed in front of the white region which denotes classical bulge. In the bins where pseudo bulge fraction dominates, classical bulge fraction is represented by the lower histogram   }
   \label{fig:stellarmass}
\end{figure}

\begin{figure*}
  \begin{subfigure}[b]{0.35\textwidth}
    \includegraphics[width=\textwidth]{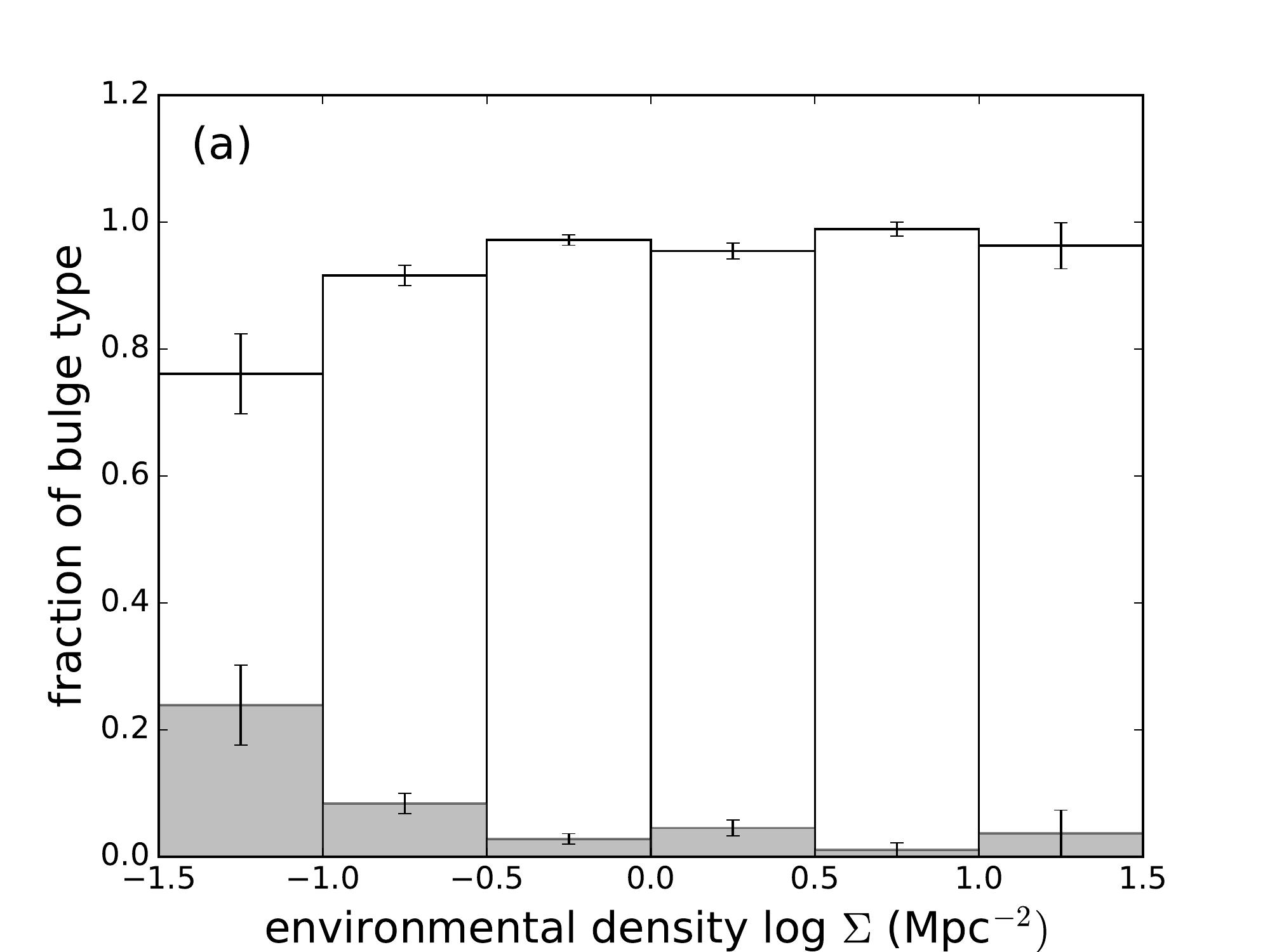}
     \phantomcaption
     \label{fig:central_sigma}
   
  \end{subfigure} 
 \hspace{-1.9em}%
  \begin{subfigure}[b]{0.35\textwidth}
    \includegraphics[width=\textwidth]{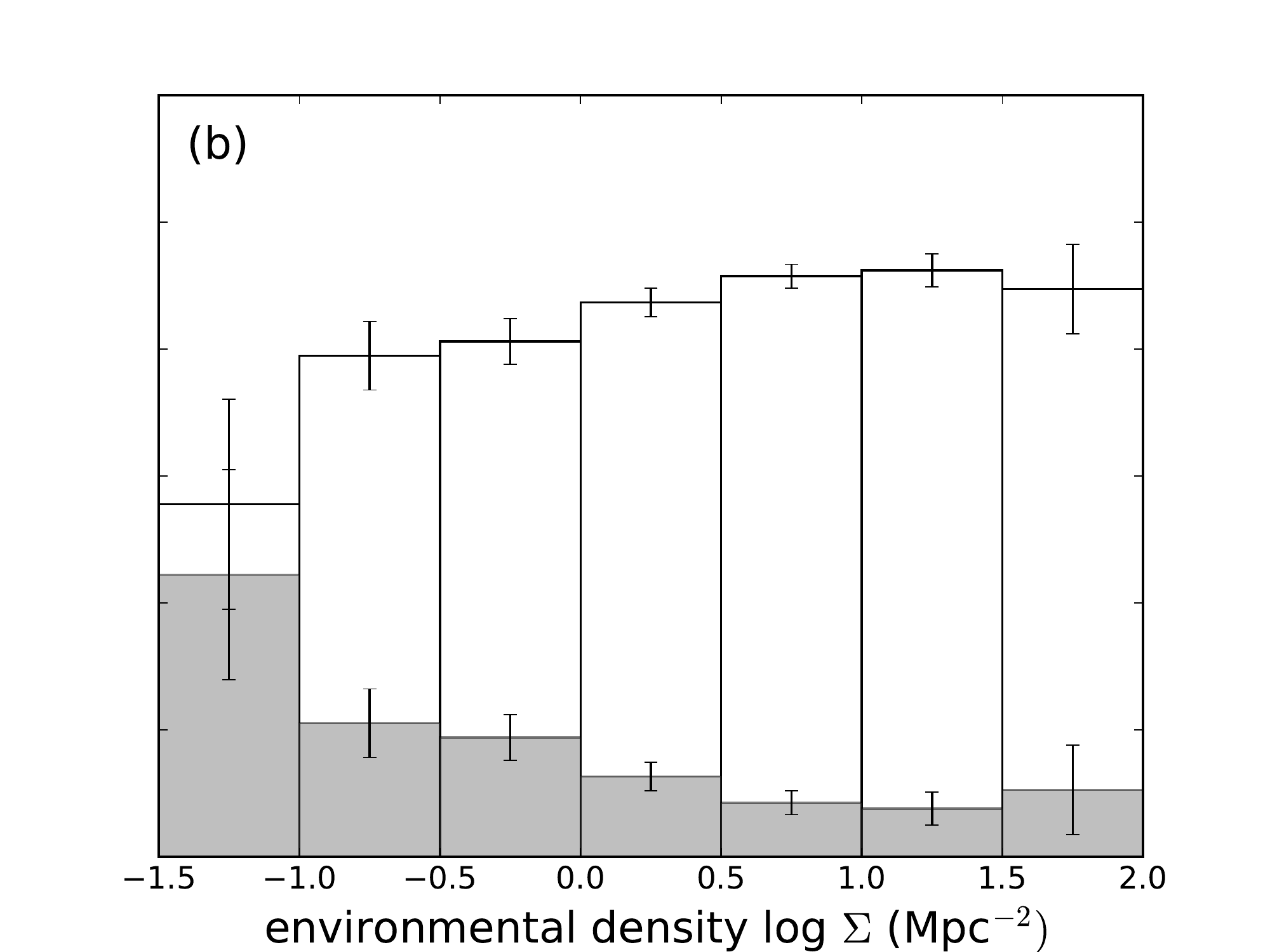}
    \phantomcaption
    \label{fig:sat_sigma}
  
  \end{subfigure} 
  \hspace{-1.9em}%
  \begin{subfigure}[b]{0.35\textwidth}
    \includegraphics[width=\textwidth]{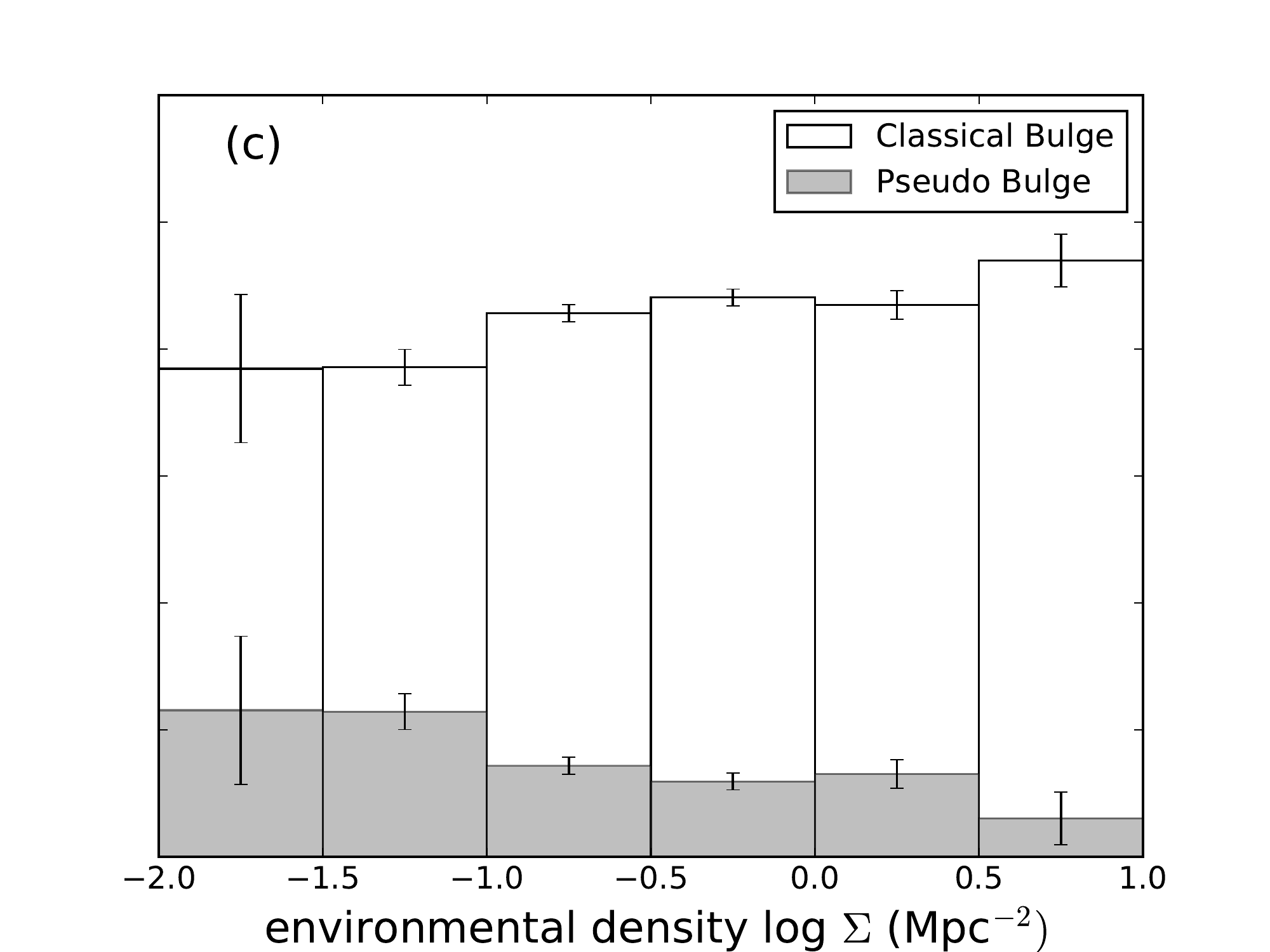}
    \phantomcaption
    \label{fig:field_sigma}
    
  \end{subfigure}
  
\caption{Dependence of fraction of bulge type as function of average environmental density for (a) central galaxies, (b) satellite galaxies and (c) field galaxies. The colour scheme is same as in Figure \ref{fig:stellarmass}.}  
\end{figure*}

\begin{figure*}
  \begin{subfigure}[b]{0.4\textwidth}
    \includegraphics[width=\textwidth]{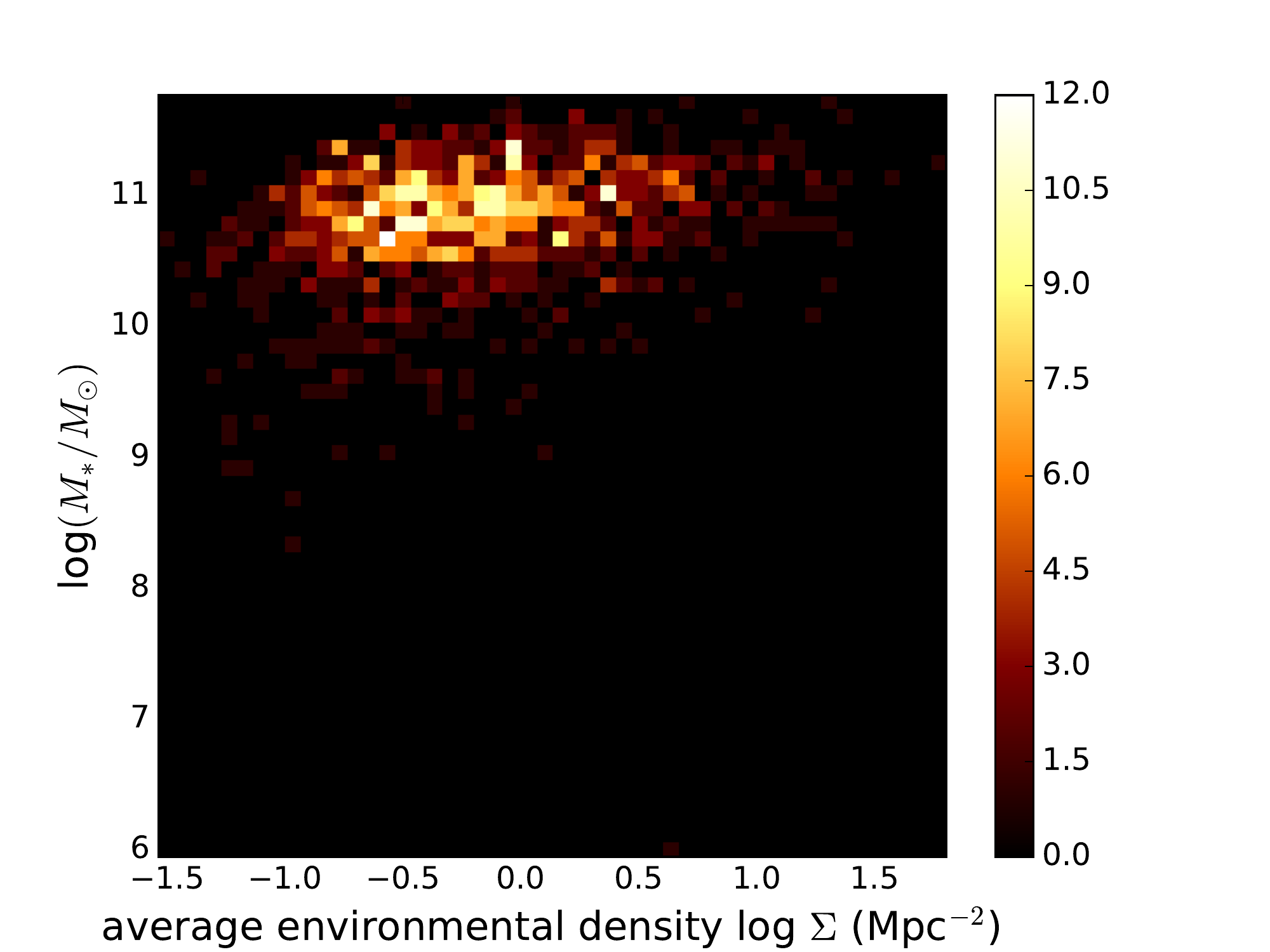}
    \caption{}
    \label{fig:2dhist_all}
    
  \end{subfigure}
  \begin{subfigure}[b]{0.4\textwidth}
    \includegraphics[width=\textwidth]{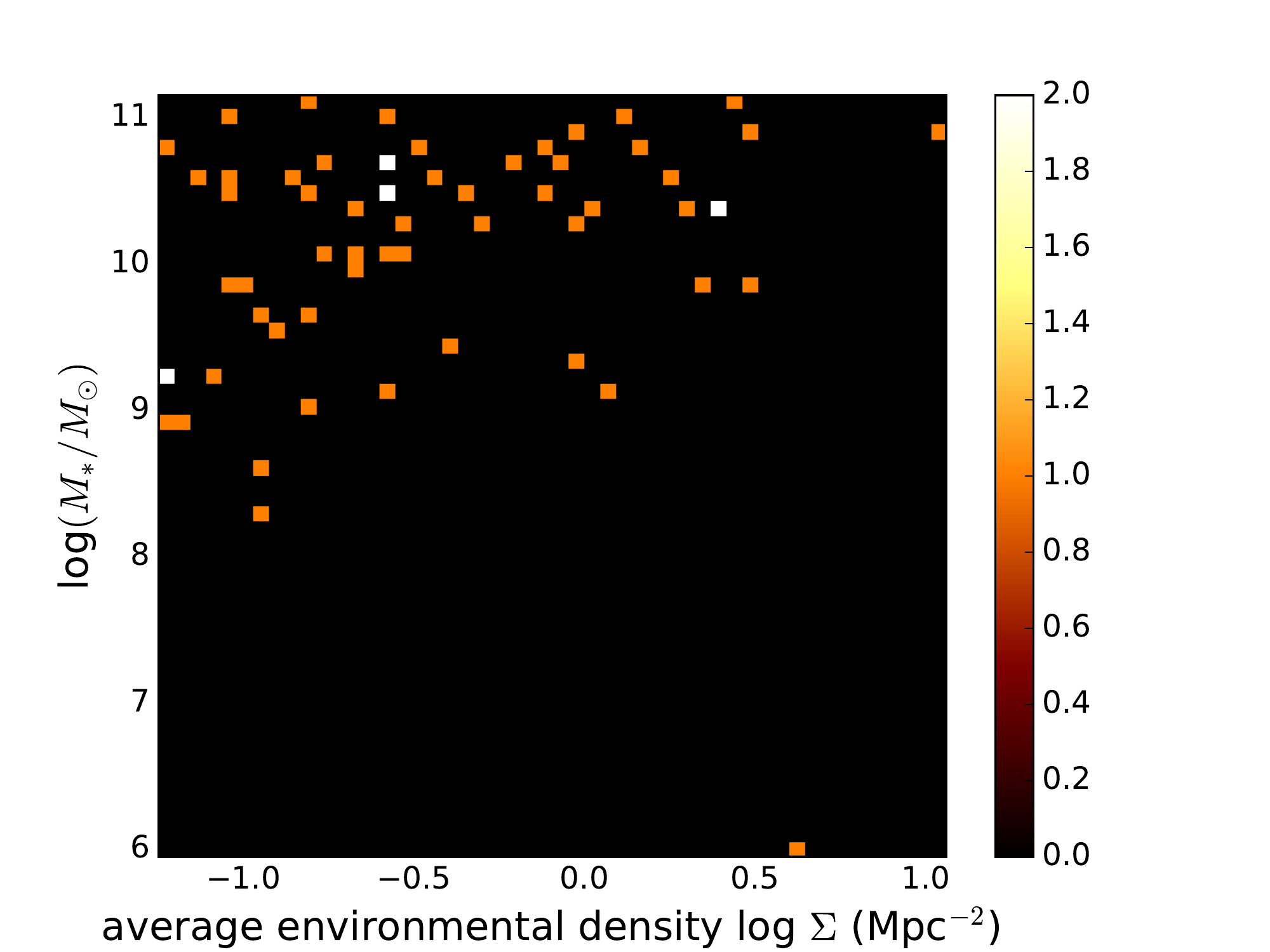}
    \caption{}
    \label{fig:2dhist_pseudo}
    
  \end{subfigure}
  
\caption{Stellar mass - average environmental density 2D histogram for (a) all central disc galaxies in our sample; (b) pseudo bulge host central disc galaxies.}  
\end{figure*}


\begin{figure*}
   \begin{subfigure}[b]{0.35\textwidth}
    \includegraphics[width=\textwidth]{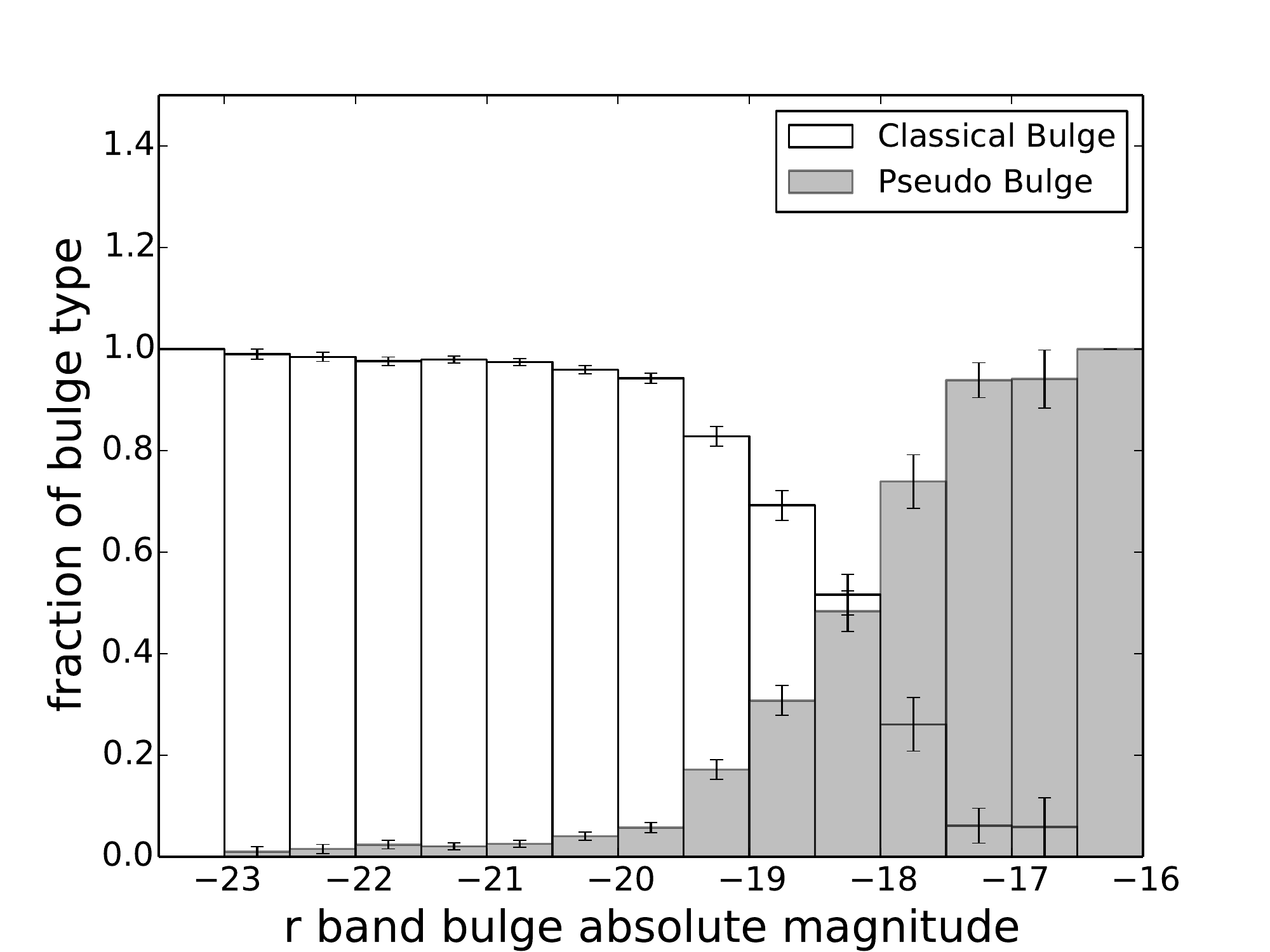}
    \caption{}
    \label{fig:absmag}
  \end{subfigure}
  \hspace{-1.7em}%
  \begin{subfigure}[b]{0.35\textwidth}
    \includegraphics[width=\textwidth]{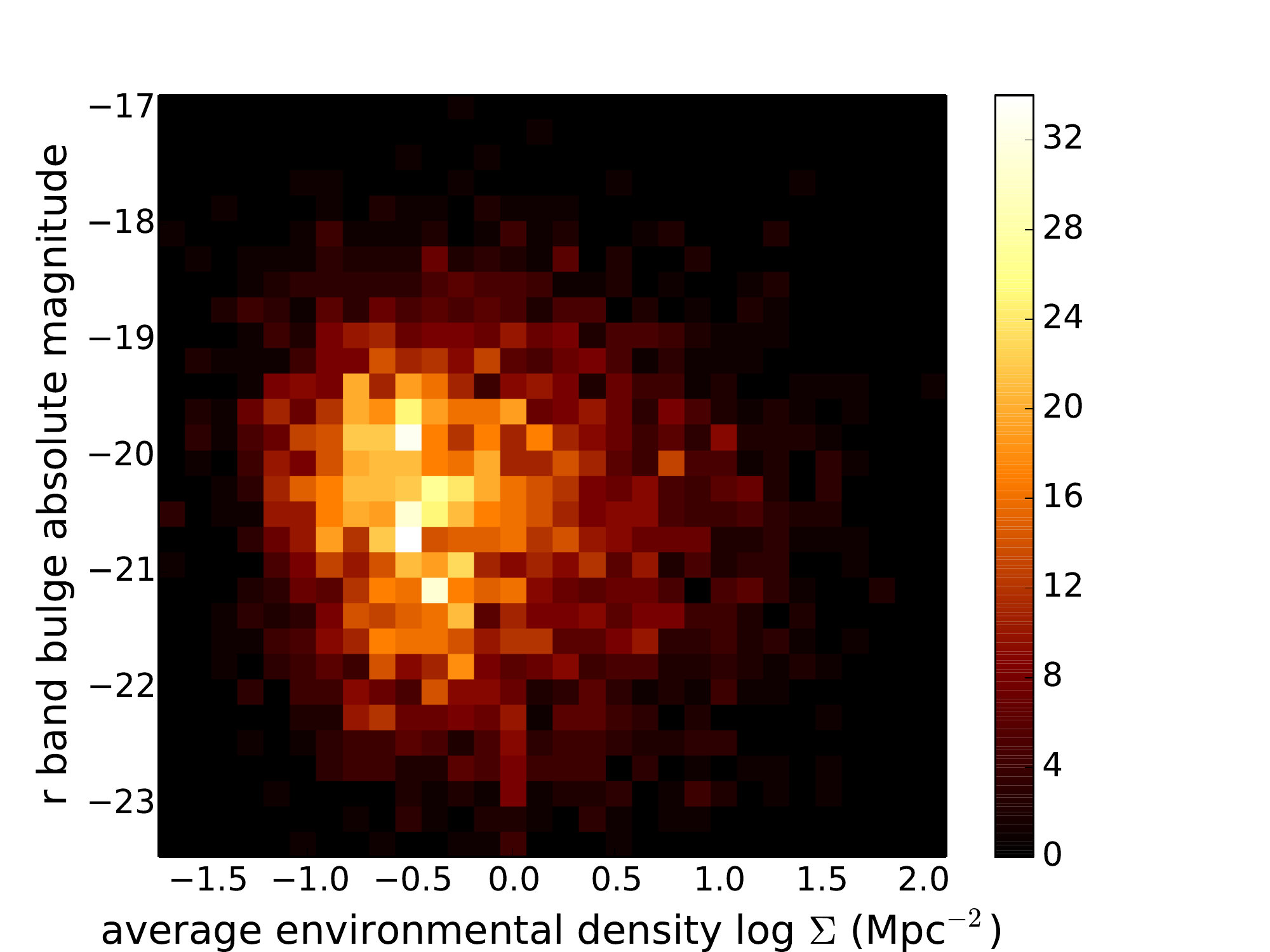}
    \caption{}
    \label{fig:2dCB_mag_sigma}
  \end{subfigure}
  \hspace{-2.5em}%
  \begin{subfigure}[b]{0.35\textwidth}
    \includegraphics[width=\textwidth]{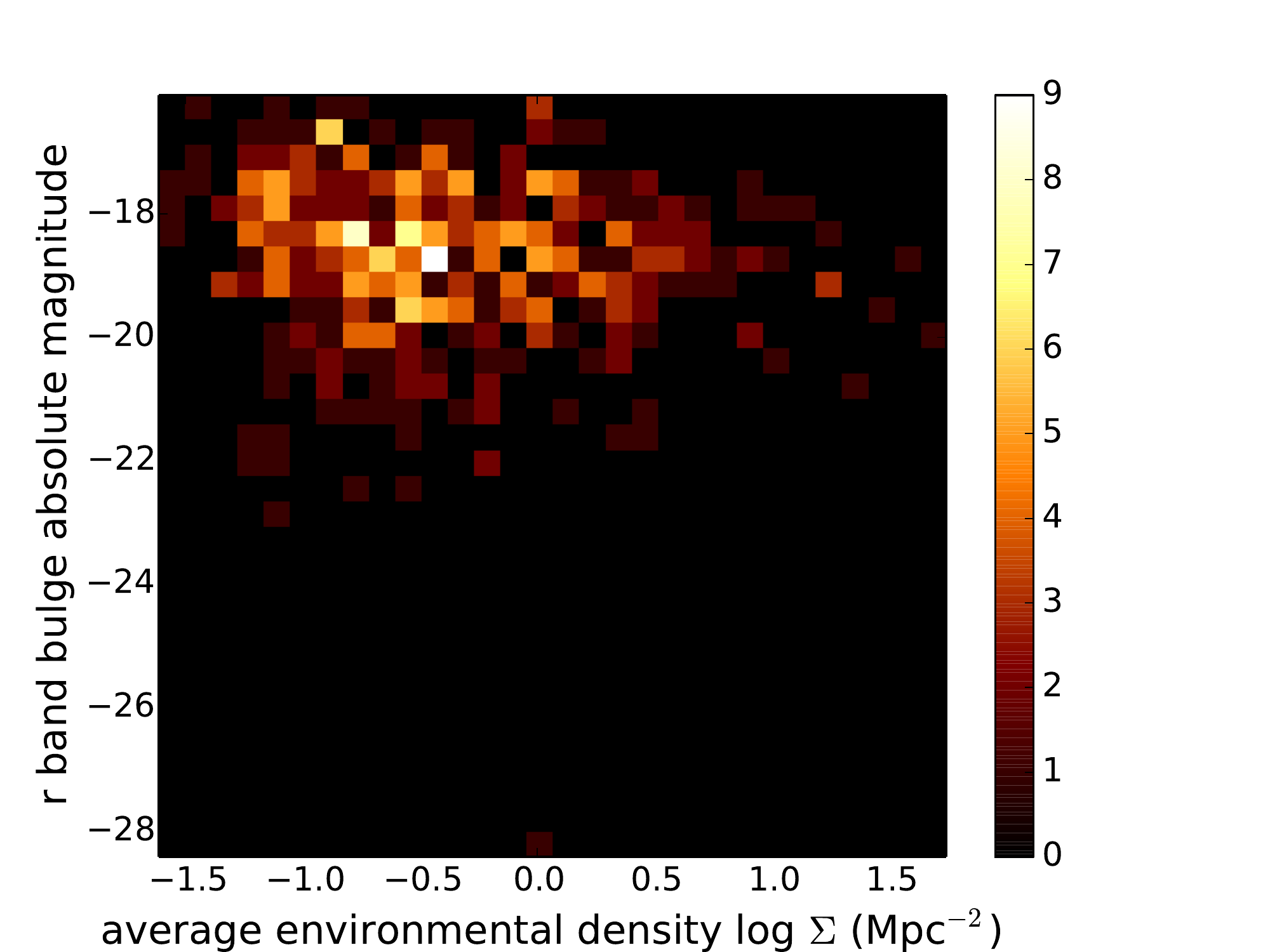}
    \caption{}
    \label{fig:2dPB_mag_sigma}
  \end{subfigure}
  \hspace{-2.5em}%
  
\label{fig:mag}

\caption{(a) Dependence of bulge fraction on bulge absolute magnitude. The colour scheme is same as in Figure \ref{fig:stellarmass}; (b) bulge absolute magnitude - average environmental density 2D histogram for all disc galaxies in our sample and (c) for pseudo bulge host disc galaxies.}  
\end{figure*}

\subsection{Bulge fraction as a function of environment}

\citet{Nair2010} provides the information of group membership and number of galaxies in a particular group or group richness taken from \citet{Yang2007}. Depending on group ID and group richness (Ngr) a flag has been provided which tells if a galaxy is the most massive member of the group or a satellite galaxy. To study the effect of environment on frequency of occurrence of bulge type, we have divided our sample in three categories. We use flags specified in \citet{Nair2010} catalogue to classify the galaxies as : \\ \\
(i) Central galaxies: are the galaxies which are most massive in a particular group and have group richness Ngr > 1. \\ \\
(ii) Satellite galaxies: are galaxies other than the central galaxy in groups with richness Ngr > 1. \\ \\
(iii) Field galaxies: are galaxies having group richness Ngr = 1. \\ \\

One should keep in mind the fact that a group as defined by \citet{Yang2007} refers to a collection of galaxies which reside in a common dark matter halo. Hence, according to this definition, clusters of galaxies having hundreds of members or just two neighbouring galaxies as long as they reside in same dark matter halo are labelled as groups. Table \ref{tab:elliptical_table} provides number distribution of ellipticals in our sample which are central, satellite or field galaxies. Tables \ref{tab:disk_all}, \ref{tab:S0_all} and \ref{tab:spiral_all} summarise the statistics of total number of galaxies specified as central, satellite and field galaxies in disc galaxies as well as in spiral and S0 galaxies separately. Comparing Tables \ref{tab:S0_all} and \ref{tab:spiral_all} we see that the pseudo bulge fraction (defined as number of pseudo bulge hosts divided by total number of galaxies) in spirals is more than 3 times of the fraction of pseudo bulge hosts in S0 galaxies. This applies to all three categories i.e. central, satellite and field galaxies of spiral and S0 morphology class. It is also interesting to note that for a specific morphology, the pseudo bulge fraction is similar for satellites and fields but becomes less than half of this value in central galaxies. \par 

\citet{Fisher2011} have reported a strong dependence of bulge type on host galaxy mass in low density environments. Since our sample spans a wide density range upto cluster environments, we checked the same dependence by plotting the fraction of bulge type across different mass bins of host galaxies. Mass of all galaxies in our sample is taken from \citet{Kauffmann2003} and  the resulting plot is shown in Figure \ref{fig:stellarmass} which shows that the pseudo bulge fraction decreases with increase in host galaxy mass while the trend is reversed for classical bulge hosts. The errors are taken as Poisson on the total number of pseudo and classical bulges and have been propagated to determine error bars on pseudo and classical bulge fractions. It is also evident from Figure \ref{fig:stellarmass} that pseudo bulge hosts dominate when host galaxy mass is less than $10^{9.5}M_{\odot}$ while classical bulges are more common above this limit. Revisiting Tables \ref{tab:disk_all}, \ref{tab:S0_all} and \ref{tab:spiral_all} with this information of stellar mass dependence of pseudo bulge hosts, we note that the median stellar mass of central, satellite and field galaxies is similar. So the fact that pseudo bulge fraction in central galaxies is half of the pseudo bulge fraction found in satellite and field galaxies seems to be an environmental effect. \par 

We now explore the dependence of bulge fraction in central, satellite and field galaxies separately on average environmental density parameter $\Sigma$ which is available in Nair catalogue taken from
\citet{Baldry2006}. It is defined as $\Sigma$=$N/(\pi d_{N}^2)$ where $d_{N}$ is projected comoving distance to the Nth nearest neighbour. \citet{Nair2010} catalogue gives a best estimate density
obtained by calculating the average density for N = 4 and N = 5 with spectroscopically confirmed members only with entire sample. For each category (central, satellite, field) of galaxies, we have divided $\Sigma$ in different bins and in each of these bins we have
calculated the fraction of galaxies hosting classical and pseudo bulges. Figures \ref{fig:central_sigma}, \ref{fig:sat_sigma} and \ref{fig:field_sigma} shows the dependence of the
fraction bulge type on average environmental density $\Sigma$ for central, satellite and disc galaxies respectively. A quick examination of these three plots tells us that within error bars pseudo bulge fraction in satellites and fields show very minor variation with respect to each other but we find a significant trend in pseudo bulge fraction with average environmental density. At lowest environmental densities pseudo bulge fraction in central galaxies is around 21\%
which steadily decreases and reaches about 5\% and becomes constant within error bars for log $ \Sigma \geq 0.0$. A point to note here is that the total number of pseudo bulges in S0 galaxies (see Table \ref{tab:S0_all}) is significantly less than total number of pseudo bulges in spiral galaxies (see Table \ref{tab:spiral_all}) in all three classes viz. central, satellite and field galaxies. As a result, the number of pseudo bulge hosting S0 galaxies will also be significantly less as compared to number of pseudo bulge hosting spirals in each bin of environmental density. Hence these trends of pseudo bulge fraction with environmental density are driven by the larger number of spiral galaxies. \par 

We need to check whether the dependence of pseudo bulge host central galaxies on environment is a direct effect or is an indirect effect induced by their common dependence on stellar mass. To do this, we have plotted a 2D histogram of galaxy stellar mass vs. average
environmental density for all central disc galaxies to check for existence of any correlation. The resultant plot is shown in Figure \ref{fig:2dhist_all} and we see no obvious dependence of stellar mass on average environmental density $\Sigma$. We checked the possibility that the high fraction of pseudo bulges as seen in the environmental density range $-1.5 < $log $\Sigma< 0.5$ in central galaxies is due to the fact that galaxies having low stellar mass might be dominating in
that density range. We have seen in Figure \ref{fig:stellarmass} that pseudo bulge hosts dominate below stellar mass $<10^{9.5}M_{\odot}$. To find out the stellar mass distribution of pseudo bulge host central galaxies across environment, we have plotted 2D histogram of stellar mass vs. average environmental density ($\Sigma$) for only those central galaxies which host a pseudo
bulge. This plot is shown in Figure \ref{fig:2dhist_pseudo} and it is clear that in the range $-1.5 <\Sigma< 0.5$ it is dominated with pseudo bulge hosts with mass $> 10^{9.5}M_{\odot}$ ruling out the possibility that for the central galaxies, high fraction of pseudo bulges seen in this density range is only a result of stellar mass dependence of pseudo bulge fraction. This result further provides support to the idea that the pseudo bulge fraction is dependent on environment. \par

To understand the influence of environment on intrinsic properties of bulge rather than their host galaxies, we have estimated the $r$ band absolute magnitude of the bulges in our sample. The dependence of bulge type on bulge absolute magnitude is shown in Figure \ref{fig:absmag} which shows that pseudo bulge fraction increases in low luminosity regime and the trend is reversed for classical bulge hosts. Figures \ref{fig:2dCB_mag_sigma} and \ref{fig:2dPB_mag_sigma} are 2D histogram plots of bulge absolute magnitude vs. average environmental density for classical and pseudo bulges in our sample respectively. Large scatter in these two plots readily tells us that
there is no strong dependence of bulge luminosity on environment.

\section{Summary \& Discussion}

We have presented a systematic study of bulges of disc galaxies in galaxy group environment. The groups are defined as galaxies that share the same dark matter halo as identified by \citet{Yang2007}. We use position of bulge on Kormendy diagram as the defining criterion for determination of bulge type. We find that 11.47\% of disc galaxies in our sample host pseudo bulges. Dividing this sample by morphology we find that 16.68\% of spiral galaxies in our sample are pseudo bulge host while this percentage is 5.37 \% for S0 galaxies. Further division of galaxies into three group based categories of central, satellite and field galaxies tells us that for the satellite and field galaxies pseudo bulge fraction is similar. On the other hand, we find that pseudo bulge fraction in central galaxies is less than half of the fraction in satellite and field galaxies, irrespective of morphology. \par
 
We find a significant dependence of pseudo bulge fraction hosted by central galaxies on average environmental density where pseudo bulges are more likely to be found in low density environments. We hardly see any such dependence of pseudo bulge fraction on environment for satellite galaxies and those which are in the field. Since galaxies having mass $< 10^{9.5}M_{\odot}$ are dominated by pseudo bulge hosts, we also have checked if the trend of pseudo bulges in central galaxies with environment is an indirect effect of stellar mass dependence of bulge type. We find that inferred dependence of pseudo bulge hosting central galaxies on environment is a likely to be direct effect of environment.\par

If pseudo bulges are formed through internal processes and evolve secularly then it seems environment plays some role in affecting the process which determines distribution of bulge type. Our finding of higher number of pseudo bulges in less dense environment is consistent with earlier studies for eg. \citep{Fisher2011} where pseudo bulges and discs were found to be more dominant in under dense void like environments as compared to galaxy clusters.  We also have found that bulge absolute magnitude which is an intrinsic property of the bulge does not depend on environmental density. Thus it might be likely that the processes which form and grow pseudo bulges are independent of environment and are governed by internal processes only.\par

However, while interpreting our results one should keep in mind the fact that bulge type is dependent on a number of parameters such as galaxy stellar mass, SFR, sSFR, morphology etc. Hence, to correctly determine the effect of environment on distribution of bulge types,
one needs to separate any effect from these parameters on which bulge type depends, as they might be contributing indirectly to some extent in the observed trends. Stellar mass and a number of parameters such as sSFR, SFR etc. are well correlated. Therefore, checking for any indirect effect of stellar mass that may show up in trend of pseudo bulges with environmental density, takes care of other quantities which are well correlated to the stellar mass. Indirect effect of other parameters such as morphology etc. on environmental dependence of pseudo bulges, have not been specifically checked in this work. \par 

Finally, we would like to remind the reader about the sample bias. From Figure \ref{fig:stellarmass} we know pseudo bulges are more commonly found in galaxies having stellar mass $< 10^{9.5}M_{\odot}$, but as shown in Figure \ref{fig:mass} number of such galaxies are very less in our sample. As a result, we have less number of pseudo bulges in our sample than one would expect in this mass range.  Therefore, the results presented in this work on pseudo bulge fraction should be understood keeping the sample bias in mind.\par 

In future, we would like to explore properties of pseudo bulges and their dependence on environment as well on morphology in greater detail. Using Galaxy Zoo data \citep{Lintott2011} which provides us with morphological information on nearly 900,000 galaxies in combination with information on group properties from \citet{Yang2007}, we will be able to explore many aspects of environmental secular evolution. Our result shows that the pseudo bulge fraction seems to be dependent on distance of galaxy from the group centre with lower fraction of pseudo bulges found in central galaxies as compared to satellites. With a large sample of galaxy groups, it will be possible to explore dependence of pseudo bulge fraction on the group centric distance of member galaxies. Dividing the group centric distance into different bins, one can also check dependence of pseudo bulge fraction on environment for the galaxies which are nearer and farther away from the galaxy group centre. Morphological information on large number of galaxies will also help us to separate dependence of pseudo bulge fraction on galaxy morphology which might be contributing to some extent on environmental dependence of pseudo bulge fraction. We believe that work in these directions will help to understand secular evolution in a quantitative way.

\section*{Acknowledgements}

We thank the anonymous referee for insightful comments that have improved both the content and presentation of this paper. PKM thanks Peter Kamphuis for useful discussions and Omkar Bait for help with Python programming. YW thanks IUCAA for hosting him on his sabbatical when this work was initiated. SB would like to acknowledge support from the National Research Foundation research grant (PID-93727). SB and YW acknowledge support from a bilateral grant under the Indo-South Africa Science and Technology Cooperation (PID-102296) funded by the Departments of Science and Technology (DST) of the Indian and South African Governments.





\bibliographystyle{mnras}
\bibliography{ref.bib}

\begin{thebibliography}{}
\makeatletter
\relax
\def\mn@urlcharsother{\let\do\@makeother \do\$\do\&\do\#\do\^\do\_\do\%\do\~}
\def\mn@doi{\begingroup\mn@urlcharsother \@ifnextchar [ {\mn@doi@}
  {\mn@doi@[]}}
\def\mn@doi@[#1]#2{\def\@tempa{#1}\ifx\@tempa\@empty \href
  {http://dx.doi.org/#2} {doi:#2}\else \href {http://dx.doi.org/#2} {#1}\fi
  \endgroup}
\def\mn@eprint#1#2{\mn@eprint@#1:#2::\@nil}
\def\mn@eprint@arXiv#1{\href {http://arxiv.org/abs/#1} {{\tt arXiv:#1}}}
\def\mn@eprint@dblp#1{\href {http://dblp.uni-trier.de/rec/bibtex/#1.xml}
  {dblp:#1}}
\def\mn@eprint@#1:#2:#3:#4\@nil{\def\@tempa {#1}\def\@tempb {#2}\def\@tempc
  {#3}\ifx \@tempc \@empty \let \@tempc \@tempb \let \@tempb \@tempa \fi \ifx
  \@tempb \@empty \def\@tempb {arXiv}\fi \@ifundefined
  {mn@eprint@\@tempb}{\@tempb:\@tempc}{\expandafter \expandafter \csname
  mn@eprint@\@tempb\endcsname \expandafter{\@tempc}}}

\bibitem[\protect\citeauthoryear{{Aguerri}, {Balcells}  \&
  {Peletier}}{{Aguerri} et~al.}{2001}]{Aguerri2001}
{Aguerri} J.~A.~L.,  {Balcells} M.,   {Peletier} R.~F.,  2001, \mn@doi [\aap]
  {10.1051/0004-6361:20000441}, \href
  {http://adsabs.harvard.edu/abs/2001A%26A...367..428A} {367, 428}

\bibitem[\protect\citeauthoryear{{Athanassoula}}{{Athanassoula}}{1992}]{Athanassoula1992}
{Athanassoula} E.,  1992, \mn@doi [\mnras] {10.1093/mnras/259.2.345}, \href
  {http://adsabs.harvard.edu/abs/1992MNRAS.259..345A} {259, 345}

\bibitem[\protect\citeauthoryear{{Baldry}, {Balogh}, {Bower}, {Glazebrook},
  {Nichol}, {Bamford}  \& {Budavari}}{{Baldry} et~al.}{2006}]{Baldry2006}
{Baldry} I.~K.,  {Balogh} M.~L.,  {Bower} R.~G.,  {Glazebrook} K.,  {Nichol}
  R.~C.,  {Bamford} S.~P.,   {Budavari} T.,  2006, \mn@doi [\mnras]
  {10.1111/j.1365-2966.2006.11081.x}, \href
  {http://adsabs.harvard.edu/abs/2006MNRAS.373..469B} {373, 469}

\bibitem[\protect\citeauthoryear{{Balogh}, {Morris}, {Yee}, {Carlberg}  \&
  {Ellingson}}{{Balogh} et~al.}{1999}]{Balogh1999}
{Balogh} M.~L.,  {Morris} S.~L.,  {Yee} H.~K.~C.,  {Carlberg} R.~G.,
  {Ellingson} E.,  1999, \mn@doi [\apj] {10.1086/308056}, \href
  {http://adsabs.harvard.edu/abs/1999ApJ...527...54B} {527, 54}

\bibitem[\protect\citeauthoryear{{Cacciato}, {Dekel}  \& {Genel}}{{Cacciato}
  et~al.}{2012}]{Cacciato2012}
{Cacciato} M.,  {Dekel} A.,   {Genel} S.,  2012, \mn@doi [\mnras]
  {10.1111/j.1365-2966.2011.20359.x}, \href
  {http://adsabs.harvard.edu/abs/2012MNRAS.421..818C} {421, 818}

\bibitem[\protect\citeauthoryear{{Carollo}, {Stiavelli}  \& {Mack}}{{Carollo}
  et~al.}{1998}]{Carollo1998}
{Carollo} C.~M.,  {Stiavelli} M.,   {Mack} J.,  1998, \mn@doi [\aj]
  {10.1086/300407}, \href {http://adsabs.harvard.edu/abs/1998AJ....116...68C}
  {116, 68}

\bibitem[\protect\citeauthoryear{{Dekel} et~al.,}{{Dekel}
  et~al.}{2009}]{Dekel2009}
{Dekel} A.,  et~al., 2009, \mn@doi [\nat] {10.1038/nature07648}, \href
  {http://adsabs.harvard.edu/abs/2009Natur.457..451D} {457, 451}

\bibitem[\protect\citeauthoryear{{Djorgovski} \& {Davis}}{{Djorgovski} \&
  {Davis}}{1987}]{Djorgovski1987}
{Djorgovski} S.,  {Davis} M.,  1987, \mn@doi [\apj] {10.1086/164948}, \href
  {http://adsabs.harvard.edu/abs/1987ApJ...313...59D} {313, 59}

\bibitem[\protect\citeauthoryear{{Durbala}, {Sulentic}, {Buta}  \&
  {Verdes-Montenegro}}{{Durbala} et~al.}{2008}]{Durbala2008}
{Durbala} A.,  {Sulentic} J.~W.,  {Buta} R.,   {Verdes-Montenegro} L.,  2008,
  \mn@doi [\mnras] {10.1111/j.1365-2966.2008.13713.x}, \href
  {http://adsabs.harvard.edu/abs/2008MNRAS.390..881D} {390, 881}

\bibitem[\protect\citeauthoryear{{Elmegreen}, {Bournaud}  \&
  {Elmegreen}}{{Elmegreen} et~al.}{2008}]{Elmegreen2008}
{Elmegreen} B.~G.,  {Bournaud} F.,   {Elmegreen} D.~M.,  2008, \mn@doi [\apj]
  {10.1086/592190}, \href {http://adsabs.harvard.edu/abs/2008ApJ...688...67E}
  {688, 67}

\bibitem[\protect\citeauthoryear{{Erwin} et~al.,}{{Erwin}
  et~al.}{2015}]{Erwin2015}
{Erwin} P.,  et~al., 2015, \mn@doi [\mnras] {10.1093/mnras/stu2376}, \href
  {http://adsabs.harvard.edu/abs/2015MNRAS.446.4039E} {446, 4039}

\bibitem[\protect\citeauthoryear{{Fabricius}, {Saglia}, {Fisher}, {Drory},
  {Bender}  \& {Hopp}}{{Fabricius} et~al.}{2012}]{Fabricius2012}
{Fabricius} M.~H.,  {Saglia} R.~P.,  {Fisher} D.~B.,  {Drory} N.,  {Bender} R.,
    {Hopp} U.,  2012, \mn@doi [\apj] {10.1088/0004-637X/754/1/67}, \href
  {http://adsabs.harvard.edu/abs/2012ApJ...754...67F} {754, 67}

\bibitem[\protect\citeauthoryear{{Fern{\'a}ndez Lorenzo}
  et~al.,}{{Fern{\'a}ndez Lorenzo} et~al.}{2014}]{Lorenzo2014}
{Fern{\'a}ndez Lorenzo} M.,  et~al., 2014, \mn@doi [\apjl]
  {10.1088/2041-8205/788/2/L39}, \href
  {http://adsabs.harvard.edu/abs/2014ApJ...788L..39F} {788, L39}

\bibitem[\protect\citeauthoryear{{Fisher}}{{Fisher}}{2006}]{Fisher2006}
{Fisher} D.~B.,  2006, \mn@doi [\apjl] {10.1086/504351}, \href
  {http://adsabs.harvard.edu/abs/2006ApJ...642L..17F} {642, L17}

\bibitem[\protect\citeauthoryear{{Fisher} \& {Drory}}{{Fisher} \&
  {Drory}}{2008}]{Fisher2008}
{Fisher} D.~B.,  {Drory} N.,  2008, \mn@doi [\aj]
  {10.1088/0004-6256/136/2/773}, \href
  {http://adsabs.harvard.edu/abs/2008AJ....136..773F} {136, 773}

\bibitem[\protect\citeauthoryear{{Fisher} \& {Drory}}{{Fisher} \&
  {Drory}}{2010}]{Fisher2010}
{Fisher} D.~B.,  {Drory} N.,  2010, \mn@doi [\apj]
  {10.1088/0004-637X/716/2/942}, \href
  {http://adsabs.harvard.edu/abs/2010ApJ...716..942F} {716, 942}

\bibitem[\protect\citeauthoryear{{Fisher} \& {Drory}}{{Fisher} \&
  {Drory}}{2011}]{Fisher2011}
{Fisher} D.~B.,  {Drory} N.,  2011, \mn@doi [\apjl]
  {10.1088/2041-8205/733/2/L47}, \href
  {http://adsabs.harvard.edu/abs/2011ApJ...733L..47F} {733, L47}

\bibitem[\protect\citeauthoryear{{Fisher} \& {Drory}}{{Fisher} \&
  {Drory}}{2016}]{Fisher2016}
{Fisher} D.~B.,  {Drory} N.,  2016, \mn@doi [Galactic Bulges]
  {10.1007/978-3-319-19378-6_3}, \href
  {http://adsabs.harvard.edu/abs/2016ASSL..418...41F} {418, 41}

\bibitem[\protect\citeauthoryear{{Forbes}, {Krumholz}, {Burkert}  \&
  {Dekel}}{{Forbes} et~al.}{2014}]{Forbes2014}
{Forbes} J.~C.,  {Krumholz} M.~R.,  {Burkert} A.,   {Dekel} A.,  2014, \mn@doi
  [\mnras] {10.1093/mnras/stt2294}, \href
  {http://adsabs.harvard.edu/abs/2014MNRAS.438.1552F} {438, 1552}

\bibitem[\protect\citeauthoryear{{Gadotti}}{{Gadotti}}{2008}]{Gadotti2008}
{Gadotti} D.~A.,  2008, \mn@doi [\mnras] {10.1111/j.1365-2966.2007.12723.x},
  \href {http://adsabs.harvard.edu/abs/2008MNRAS.384..420G} {384, 420}

\bibitem[\protect\citeauthoryear{{Gadotti}}{{Gadotti}}{2009}]{Gadotti2009}
{Gadotti} D.~A.,  2009, \mn@doi [\mnras] {10.1111/j.1365-2966.2008.14257.x},
  \href {http://adsabs.harvard.edu/abs/2009MNRAS.393.1531G} {393, 1531}

\bibitem[\protect\citeauthoryear{{Gadotti} \& {Kauffmann}}{{Gadotti} \&
  {Kauffmann}}{2009}]{Gadotti2009a}
{Gadotti} D.~A.,  {Kauffmann} G.,  2009, \mn@doi [\mnras]
  {10.1111/j.1365-2966.2009.15328.x}, \href
  {http://adsabs.harvard.edu/abs/2009MNRAS.399..621G} {399, 621}

\bibitem[\protect\citeauthoryear{{Gadotti} \& {dos Anjos}}{{Gadotti} \& {dos
  Anjos}}{2001}]{Gadotti2001}
{Gadotti} D.~A.,  {dos Anjos} S.,  2001, \mn@doi [\aj] {10.1086/322126}, \href
  {http://adsabs.harvard.edu/abs/2001AJ....122.1298G} {122, 1298}

\bibitem[\protect\citeauthoryear{{Ho} \& {Kim}}{{Ho} \& {Kim}}{2014}]{Ho2014}
{Ho} L.~C.,  {Kim} M.,  2014, \mn@doi [\apj] {10.1088/0004-637X/789/1/17},
  \href {http://adsabs.harvard.edu/abs/2014ApJ...789...17H} {789, 17}

\bibitem[\protect\citeauthoryear{{Hudson}, {Stevenson}, {Smith}, {Wegner},
  {Lucey}  \& {Simard}}{{Hudson} et~al.}{2010}]{Hudson2010}
{Hudson} M.~J.,  {Stevenson} J.~B.,  {Smith} R.~J.,  {Wegner} G.~A.,  {Lucey}
  J.~R.,   {Simard} L.,  2010, \mn@doi [\mnras]
  {10.1111/j.1365-2966.2010.17318.x}, \href
  {http://adsabs.harvard.edu/abs/2010MNRAS.409..405H} {409, 405}

\bibitem[\protect\citeauthoryear{{Kauffmann} et~al.,}{{Kauffmann}
  et~al.}{2003}]{Kauffmann2003}
{Kauffmann} G.,  et~al., 2003, \mn@doi [\mnras]
  {10.1046/j.1365-8711.2003.06291.x}, \href
  {http://adsabs.harvard.edu/abs/2003MNRAS.341...33K} {341, 33}

\bibitem[\protect\citeauthoryear{{Kormendy}}{{Kormendy}}{1977}]{Kormendy1977}
{Kormendy} J.,  1977, \mn@doi [\apj] {10.1086/155687}, \href
  {http://adsabs.harvard.edu/abs/1977ApJ...218..333K} {218, 333}

\bibitem[\protect\citeauthoryear{{Kormendy}}{{Kormendy}}{2016}]{Kormendy2016}
{Kormendy} J.,  2016, \mn@doi [Galactic Bulges] {10.1007/978-3-319-19378-6_16},
  \href {http://adsabs.harvard.edu/abs/2016ASSL..418..431K} {418, 431}

\bibitem[\protect\citeauthoryear{{Kormendy} \& {Kennicutt}}{{Kormendy} \&
  {Kennicutt}}{2004}]{Kormendy&Kennicutt2004}
{Kormendy} J.,  {Kennicutt} Jr. R.~C.,  2004, \mn@doi [\araa]
  {10.1146/annurev.astro.42.053102.134024}, \href
  {http://adsabs.harvard.edu/abs/2004ARA%26A..42..603K} {42, 603}

\bibitem[\protect\citeauthoryear{{Kormendy}, {Fisher}, {Cornell}  \&
  {Bender}}{{Kormendy} et~al.}{2009}]{Kormendy2009}
{Kormendy} J.,  {Fisher} D.~B.,  {Cornell} M.~E.,   {Bender} R.,  2009, \mn@doi
  [\apjs] {10.1088/0067-0049/182/1/216}, \href
  {http://adsabs.harvard.edu/abs/2009ApJS..182..216K} {182, 216}

\bibitem[\protect\citeauthoryear{{Kormendy}, {Bender}  \& {Cornell}}{{Kormendy}
  et~al.}{2011}]{Kormendy2011}
{Kormendy} J.,  {Bender} R.,   {Cornell} M.~E.,  2011, \mn@doi [\nat]
  {10.1038/nature09694}, \href
  {http://adsabs.harvard.edu/abs/2011Natur.469..374K} {469, 374}

\bibitem[\protect\citeauthoryear{{Lackner} \& {Gunn}}{{Lackner} \&
  {Gunn}}{2013}]{Lackner2013}
{Lackner} C.~N.,  {Gunn} J.~E.,  2013, \mn@doi [\mnras] {10.1093/mnras/sts179},
  \href {http://adsabs.harvard.edu/abs/2013MNRAS.428.2141L} {428, 2141}

\bibitem[\protect\citeauthoryear{{Lintott} et~al.,}{{Lintott}
  et~al.}{2011}]{Lintott2011}
{Lintott} C.,  et~al., 2011, \mn@doi [\mnras]
  {10.1111/j.1365-2966.2010.17432.x}, \href
  {http://adsabs.harvard.edu/abs/2011MNRAS.410..166L} {410, 166}

\bibitem[\protect\citeauthoryear{{Meert}, {Vikram}  \& {Bernardi}}{{Meert}
  et~al.}{2015}]{Meert2015}
{Meert} A.,  {Vikram} V.,   {Bernardi} M.,  2015, \mn@doi [\mnras]
  {10.1093/mnras/stu2333}, \href
  {http://adsabs.harvard.edu/abs/2015MNRAS.446.3943M} {446, 3943}

\bibitem[\protect\citeauthoryear{{Meert}, {Vikram}  \& {Bernardi}}{{Meert}
  et~al.}{2016}]{Meert2016}
{Meert} A.,  {Vikram} V.,   {Bernardi} M.,  2016, \mn@doi [\mnras]
  {10.1093/mnras/stv2475}, \href
  {http://adsabs.harvard.edu/abs/2016MNRAS.455.2440M} {455, 2440}

\bibitem[\protect\citeauthoryear{{Nair} \& {Abraham}}{{Nair} \&
  {Abraham}}{2010}]{Nair2010}
{Nair} P.~B.,  {Abraham} R.~G.,  2010, \mn@doi [\apjs]
  {10.1088/0067-0049/186/2/427}, \href
  {http://adsabs.harvard.edu/abs/2010ApJS..186..427N} {186, 427}

\bibitem[\protect\citeauthoryear{{Rautiainen} \& {Salo}}{{Rautiainen} \&
  {Salo}}{2000}]{Rautiainen2000}
{Rautiainen} P.,  {Salo} H.,  2000, \aap, \href
  {http://adsabs.harvard.edu/abs/2000A%26A...362..465R} {362, 465}

\bibitem[\protect\citeauthoryear{{Ribeiro}, {Lobo}, {Ant{\'o}n}, {Gomes}  \&
  {Papaderos}}{{Ribeiro} et~al.}{2016}]{Ribeiro2016}
{Ribeiro} B.,  {Lobo} C.,  {Ant{\'o}n} S.,  {Gomes} J.~M.,   {Papaderos} P.,
  2016, \mn@doi [\mnras] {10.1093/mnras/stv2872}, \href
  {http://adsabs.harvard.edu/abs/2016MNRAS.456.3899R} {456, 3899}

\bibitem[\protect\citeauthoryear{{Salo}, {Rautiainen}, {Buta}, {Purcell},
  {Cobb}, {Crocker}  \& {Laurikainen}}{{Salo} et~al.}{1999}]{Salo1999}
{Salo} H.,  {Rautiainen} P.,  {Buta} R.,  {Purcell} G.~B.,  {Cobb} M.~L.,
  {Crocker} D.~A.,   {Laurikainen} E.,  1999, \mn@doi [\aj] {10.1086/300726},
  \href {http://adsabs.harvard.edu/abs/1999AJ....117..792S} {117, 792}

\bibitem[\protect\citeauthoryear{{Sanders} \& {Tubbs}}{{Sanders} \&
  {Tubbs}}{1980}]{Sanders1980}
{Sanders} R.~H.,  {Tubbs} A.~D.,  1980, \mn@doi [\apj] {10.1086/157683}, \href
  {http://adsabs.harvard.edu/abs/1980ApJ...235..803S} {235, 803}

\bibitem[\protect\citeauthoryear{{Simkin}, {Su}  \& {Schwarz}}{{Simkin}
  et~al.}{1980}]{Simkin1980}
{Simkin} S.~M.,  {Su} H.~J.,   {Schwarz} M.~P.,  1980, \mn@doi [\apj]
  {10.1086/157882}, \href {http://adsabs.harvard.edu/abs/1980ApJ...237..404S}
  {237, 404}

\bibitem[\protect\citeauthoryear{{Vaghmare}, {Barway}  \&
  {Kembhavi}}{{Vaghmare} et~al.}{2013}]{Vaghmare2013}
{Vaghmare} K.,  {Barway} S.,   {Kembhavi} A.,  2013, \mn@doi [\apjl]
  {10.1088/2041-8205/767/2/L33}, \href
  {http://adsabs.harvard.edu/abs/2013ApJ...767L..33V} {767, L33}

\bibitem[\protect\citeauthoryear{{Vaghmare}, {Barway}, {Mathur}  \&
  {Kembhavi}}{{Vaghmare} et~al.}{2015}]{Vaghmare2015}
{Vaghmare} K.,  {Barway} S.,  {Mathur} S.,   {Kembhavi} A.~K.,  2015, \mn@doi
  [\mnras] {10.1093/mnras/stv668}, \href
  {http://adsabs.harvard.edu/abs/2015MNRAS.450..873V} {450, 873}

\bibitem[\protect\citeauthoryear{{Vikram}, {Wadadekar}, {Kembhavi}  \&
  {Vijayagovindan}}{{Vikram} et~al.}{2010}]{Vikram2010}
{Vikram} V.,  {Wadadekar} Y.,  {Kembhavi} A.~K.,   {Vijayagovindan} G.~V.,
  2010, \mn@doi [\mnras] {10.1111/j.1365-2966.2010.17426.x}, \href
  {http://adsabs.harvard.edu/abs/2010MNRAS.409.1379V} {409, 1379}

\bibitem[\protect\citeauthoryear{{Yang}, {Mo}, {van den Bosch}, {Pasquali},
  {Li}  \& {Barden}}{{Yang} et~al.}{2007}]{Yang2007}
{Yang} X.,  {Mo} H.~J.,  {van den Bosch} F.~C.,  {Pasquali} A.,  {Li} C.,
  {Barden} M.,  2007, \mn@doi [\apj] {10.1086/522027}, \href
  {http://adsabs.harvard.edu/abs/2007ApJ...671..153Y} {671, 153}

\makeatother
\end{thebibliography}







\bsp	
\label{lastpage}
\end{document}